\documentclass[useAMS,usenatbib]{mn2e}
\def\Msun{\hbox{M$_\odot$}}
\def\Msol{\hbox{M$_\odot$}}

\def\kms{\hbox{km$\,$s$^{-1}$}}
\def\cm3{\hbox{cm$^{-3}$}}
\def\Apix{\hbox{\AA$\,$pix$^{-1}$}}
\newcommand{\Zsol}{\ensuremath{Z_{\odot}}}
\newcommand\fsec{\hbox{$.\!\!^{\rm s}$}}
\def\one{\,{\sc i}}             % for producing Na I as Na\one\ etc.
\def\two{\,{\sc ii}}
\def\three{\,{\sc iii}}
\def\four{\,{\sc iv}}

\def\six{\,{\sc vi}}

\usepackage{upgreek}
\usepackage{color}
\usepackage{graphicx}
\usepackage{lscape}
\usepackage{rotating}

\defcitealias{westm07b}{Paper~II}
\title[A young star cluster's environment in NGC 1569]{Gemini GMOS/IFU spectroscopy of NGC 1569 -- I: Mapping the properties of a young star cluster and its environment}
\author[M.S. Westmoquette et al.] {M.S. Westmoquette$^1$\thanks{E-mail: msw@star.ucl.ac.uk}, K.M. Exter$^{2,}$\thanks{Current address: Space Telescope Science Institute, 3700 San Martin Drive, Baltimore, MD 21218, USA}, L. J. Smith$^{1,3}$ and J. S. Gallagher III$^4$\\
$^1$Department of Physics and Astronomy, University College London, Gower Street, London, WC1E 6BT\\
$^2$Instituto de Astrof\'isca de Canarias, C/Via Lactea s/n, E38200 - La Laguna (Tenerife), Espa\~ na\\
$^3$Space Telescope Science Institute and European Space Agency, 3700 San Martin Drive, Baltimore, MD 21218, USA\\
$^4$Department of Astronomy, University of Wisconsin-Madison, 5534 Sterling, 475 North Charter St., Madison WI 53706, USA\\
}
\date{Accepted 2007 August 3. Received 2007 July 12; in original form 2007 May 31}
\pagerange{\pageref{firstpage}--\pageref{lastpage}}
\pubyear{2007}
\begin{document}
\maketitle
\label{firstpage}
%%%%%%%%%%%%%%%%%%%%%%%%%%%%
\begin{abstract}
We present Gemini-North GMOS/IFU observations of a young star cluster and its environment near the centre of the dwarf irregular starburst galaxy NGC 1569. This forms part of a larger and on-going study of the formation and collimation mechanisms of galactic winds, including three additional IFU pointings in NGC 1569 covering the base of the galactic wind which are analysed in a companion paper. The good spatial- and spectral-resolution of these GMOS/IFU observations, covering 4740--6860~\AA, allow us to probe the interactions between clusters and their environments on small scales. For cluster 10, we combine the GMOS spectrum with \textit{HST} imaging to derive its properties. We find that it is composed of two very close components with ages of 5--7~Myr and $\le$5~Myr, and a combined mass of $7\pm5 \times 10^{3}$~\Msun. A strong red Wolf-Rayet (WR) emission feature confirms our young derived cluster ages.

A detailed analysis of the H$\alpha$ emission line profile shapes across the whole field-of-view shows them to be composed of a bright narrow feature (intrinsic FWHM $\sim50$~\kms) superimposed on a fainter broad component (FWHM $\leq300$~\kms). By mapping the properties of each individual component, we investigate the small-scale structure and properties of the ionized ISM, including reddening, excitation and electron densities, and for the first time find spatial correlations between the line component properties. We discuss in detail the possible mechanisms that could give rise to the two components and these correlations, and conclude that the most likely explanation for the broad emission is that it is produced in a turbulent mixing layer on the surface of the cool gas clumps embedded within the hot, fast-flowing cluster winds. We discuss implications for the mass-loading of the flow under these circumstances. The average radial velocity difference between the narrow and broad components is small compared to the line widths, implying that within the IFU field-of-view, turbulent motions dominate over large-scale bulk motions. We are therefore sampling well within the outer bounding shocks of the expanding superbubbles and within the outflow `energy injection zone'.
\end{abstract}


%%%%%%%%%%%%%%%%%%%%%%%%%%%%
\section{Introduction}\label{intro}

This is the first in a series of papers presenting integral field unit (IFU) observations of ionized gas in the dwarf starburst galaxy NGC 1569. The study is aimed at examining the small-scale relationship between the interstellar medium (ISM) and star-cluster population. In this paper we focus on the data reduction techniques and present the first results.

At a distance of $2.2\pm0.6$~Mpc \citep{israel88}, NGC 1569 is one of the closest examples of a starburst and an excellent analogue to high-redshift dwarf starbursts implicated in theories of galaxy formation. The total $B$ magnitude of NGC 1569, $M_{\rm B}=-17$ \citep{israel88}, and a remarkably homogeneous metallicity of $0.25$\,\Zsol\ (\citealt*{devost97}, \citealt{kobulnicky97}) makes this galaxy similar (in some ways) to the Small Magellanic Cloud. 

NGC 1569 has attracted considerable attention because observations at a variety of wavelengths show that it has recently undergone a galaxy-wide burst of star formation that peaked between 100~Myr and 5--10 Myr ago \citep{greggio98}. The presence of numerous H\two\ regions reveals that there is still substantial ongoing star formation \citep{waller91}. The two well-known super star clusters (SSCs) A and B \citep{ables71, arp85} are prominent products of the starburst episode, which may have been triggered by an interaction with a nearby H\one\ cloud \citep{stil98, muhle05}. An extended system of H$\alpha$ filaments, indicative of a bipolar outflow is seen \citep{heckman95, martin98}. X-ray observations show that the wind is metal-enriched and emanates from the full extent of the H$\alpha$ disc \citep*{martin02}, but only from $\sim$0.6 times the extent of the H\one\ disc \citep[][their fig.~7]{waller91, stil02, martin02}.

The star-cluster population of NGC 1569 has been studied extensively using \textit{HST} observations. \citet{hunter00} identified and catalogued a total of 48 compact but resolved clusters using WFPC2 imaging. Using the same data but different criteria, \citet{anders04} found 179 clusters, with the majority being formed in an intense starburst event which began $\sim$25 Myr ago. Of the cluster population, only the most luminous clusters A, B and No.\ 30 \citep[nomenclature from][]{hunter00} have been studied in any detail. These three SSCs alone provide a significant fraction (20--25 per cent) of the total optical and near-infrared light in the central region of NGC 1569 \citep{origlia01}. Cluster A has actually been resolved into two clusters \citep{demarchi97}, and has an integrated age of $\sim$7~Myr, with a probable small age difference between the two components \citep{hunter00, maoz01, origlia01}. Clusters B and 30 are older with ages of 10--20 and $\sim$30--50~Myr respectively \citep{hunter00, origlia01, anders04}.

The most conspicuous active star-forming region is located 90 pc to the west of cluster A and corresponds to the brightest H\two\ region in NGC 1569 \citep[No.\ 2;][]{waller91}, and the peak of the thermal radio emission \citep{lisenfeld04}, the dust emission \citep{lisenfeld02} and the sub-mm/FIR emission \citep{galliano03}. This region is at the eastern edge of a large CO cloud complex \citep{taylor99}, which contains and is surrounded by a number of radio continuum sources \citep{greve02}. \citet{hunter00} find two clusters in this region, numbers 6 and 10, the latter of which is found by \citet{tokura06} to be surrounded on one side by mid-IR [S\four]10.5\,$\upmu$m, emission coincident with a gas knot clearly visible in H$\alpha$. Cluster 10, one of the subjects of this paper, is the third visually brightest cluster after A and B \citep{hunter00}, but has not been studied in any detail thus far.

Stellar-wind and supernova-driven outflows powered by the collective injection of kinetic energy and momentum from massive stars in starbursts can drastically affect the structure and subsequent evolution of dwarf galaxies. In NGC 1569, H$\alpha$ images show a chaotic, complex, ionized structure with filaments, bubbles and loops (\citealt*{hunter93, tomita94}; \citealt{heckman95, anders04}) while the neutral ISM is highly turbulent \citep{stil02}. \citet{heckman95} observe H$\alpha$ line profiles which are ``distinctly non-Gaussian'', exhibiting broad, asymmetric wings that they decompose using multi-Gaussian profile fitting. Within $2''$ of SSC A, the H$\alpha$ line has a full-width zero-intensity of 30--50~\AA\, and they find the wings to account for up to $\sim$30 per cent of the total H$\alpha$ flux within this region. Broad emission line wings have been detected in many giant H\two{} regions both in nearby galaxies (30 Dor: \citealt{chuken94}; \citealt*{melnick99}; NGC 604: \citealt{yang96}; NGC 2363: \citealt{roy92, g-d94}) and in more distant dwarf galaxies (\citealt{izotov96, homeier99, marlowe95}; \citealt*{sidoli06}). \nocite{westm07c}Westmoquette et al.\ (2007b, in prep.) present high-resolution \textit{HST}/STIS observations of the central regions of M82, and identify a ubiquitous underlying broad component to the optical nebular emission lines in this galaxy. They were able to track the properties of this component through the central starburst regions, and relate them to the highly fragmented state of the ISM.

The nature of the energy source for these broad lines is not clear, and a number of contesting explanations have been proposed, all of which relate to the action of stellar winds and supernovae (SNe): broad lines resulting from the hot, turbulent gas confined to large bubbles blown by the winds; large velocities associated with supernova remnants (SNRs); integrating over many ionized structures (shells) with different expansion velocities; or from blow-out phenomena such as champagne flows or superwinds. These mechanisms act to differing degrees over differing spatial extents depending on the ambient physical conditions; the line widths and integrated shapes thus change accordingly. This makes an assessment of their effect or importance difficult, particularly since the available observations are often conflicting. Through the data presented here, we attempt to clarify which mechanisms could apply in the case of NGC 1569.

{\it HST} narrow-band H$\alpha$ images, in particular, illustrate that the ISM contains many small-scale structures \citep[e.g.][]{buckalew06}. To probe the roots of the wind outflow from NGC 1569, it is necessary to investigate the interaction of the individual winds from clusters with their environments on high angular-scales. Good spectral-resolution and spatial coverage are, however, equally important for determining properties such as gas dynamics and excitation mechanisms. Integral field spectroscopy (IFS) is ideally suited to these type of observations, and modern, large format integral field units (IFUs), such as those found on 8m-class telescopes, currently provide the best opportunity to fulfill these demanding requirements. Historically, it has been very difficult to manage the data products from such instruments, but in recent years the infrastructure to deal with the reduction, analysis and visualisation of IFS data has improved dramatically. 

To this end we have obtained Gemini-North GMOS/IFU observations of the cluster wind--ISM interaction zone in the central region of NGC 1569. Our observations focus on the small-scale and are of high enough spatial-resolution to study the wind--ISM interaction processes in detail. In this first paper we present the dataset for one of four IFU positions, and describe in detail the reduction and analysis techniques employed and the software tools we have developed. In a second paper \nocite{westm07b}(Westmoquette et al. 2007a; Paper II) we present the analysis of the remaining IFU positions covering the central outflow region. In a third paper (Westmoquette et al., in prep.; Paper III) we will examine the properties of the outer wind regions using deep H$\alpha$ imaging and IFU observations.

%%%%%%%%%%%%%%%%%%%%%%%%%%%%%%%%
\section{Observations and Data Reduction}\label{data}

\subsection{Observations}\label{obs}
In November 2004 queue-mode observations using the Gemini-North Multi-Object Spectrograph (GMOS) Integral Field Unit \citep[IFU;][]{allington02} were obtained covering four regions near the centre of NGC 1569 (programme ID: GN-2004B-Q-33, PI: L.J.\ Smith), with 0.5--0.8~arcsec seeing (6--9~pc at the distance of NGC 1569). A nearby bright star was used to provide guiding and tip-tilt corrections using the GMOS on-instrument wave front sensor (OIWFS). 

Depending on the combination of spatial and spectral coverage required, the GMOS IFU can be operated in one- or two-slit modes. For our purposes, we opted for the one-slit mode giving a field-of-view of $5\times 3.5$~arcsecs (which corresponds to approximately \mbox{$50\times 35$~pc} at the distance of NGC 1569) sampled by 500 hexagonal contiguous fibres of $0\farcs2$ diameter. An additional block of 250 fibres (covering $2.5\times 1.7$~arcsecs) are offset by $1'$ from the object field providing a dedicated sky view. The coordinates of the IFU field centre analysed in this paper were $\alpha =04^{\rm h}\,30^{\rm m}\,47\fsec27$; $\delta = +64^\circ\,51'\,03\farcs2$ (J2000) and the total exposure time, split into four separate integrations, was 6720~s. Using the R831 grating to give a spectral coverage of 4740--6860~\AA{} and a dispersion of 0.34~\Apix, we were able to cover the important nebular diagnostic lines of H$\alpha$, H$\beta$, [O\three]$\lambda\lambda 4959,5007$, [N\two]$\lambda\lambda 6548,6583$ and [S\two]$\lambda\lambda 6716,6731$.

The GMOS spectrograph is fed by optical fibres from the IFU which reformats the arrangement of the spectra for imaging by the detector. This is composed of three 2048\,$\times$\,4068 CCD chips separated by small gaps. In order to remove the pixel-to-pixel sensitivity differences, and enable the wavelength and flux calibration of the data, a number of bias frames, flat-fields, twilight flats, arc calibration frames, and observations of the photometric standard star G191-B2B, were taken contemporaneously with the science fields.

\begin{figure*}
\centering
\includegraphics[width=15cm]{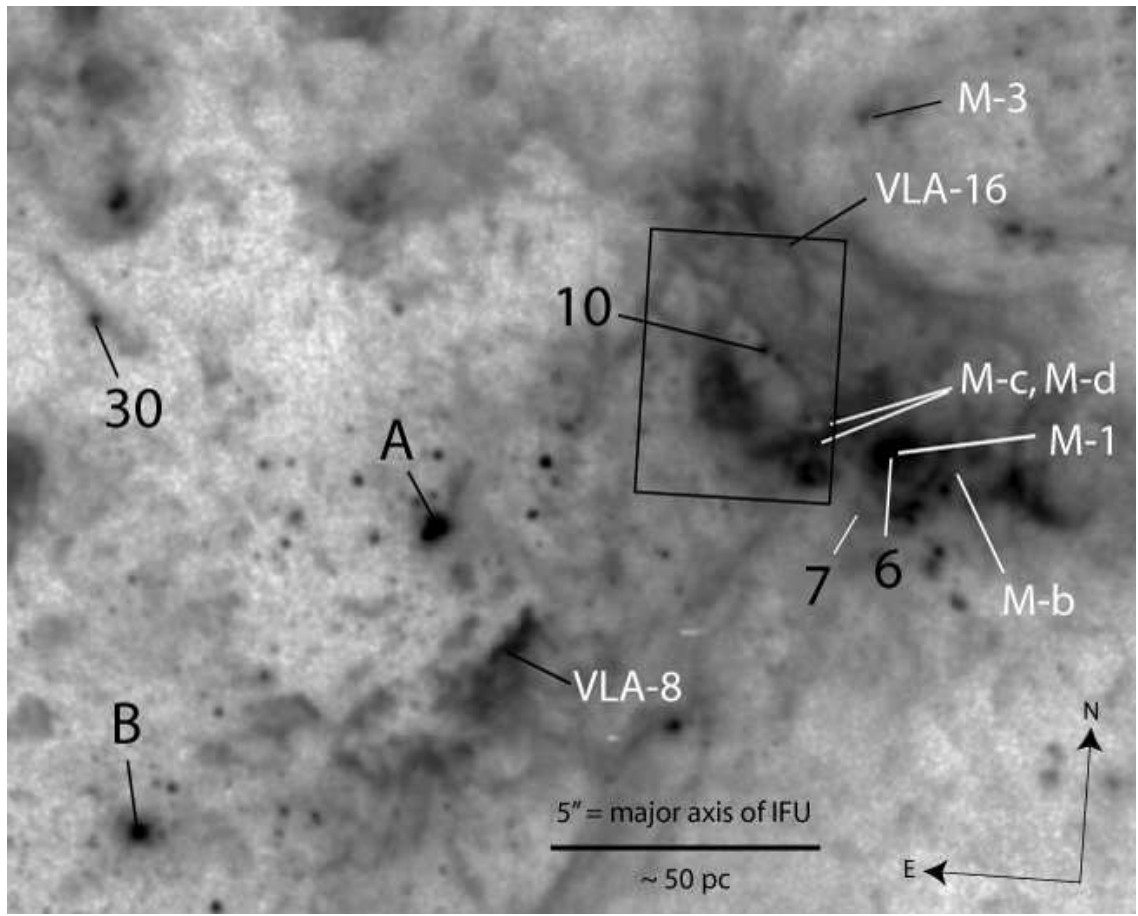}
\caption{{\it HST}/WFPC2 Planetary Camera image taken through the F656N filter of the central region of NGC 1569 showing the position of the IFU field. A number of the most prominent star clusters \citep{hunter00} and radio continuum sources \citep{greve02} are labelled in black and white respectively.}
%1'' = 10.7 pc
\label{fig:finder}
\end{figure*}

The IFU positions were chosen to cover selected areas of the disturbed ionized interstellar medium in the inner region of NGC 1569, near the roots of the wind outflow. In this paper we present data for one position which samples cluster 10 and its immediate environment. In Fig.~\ref{fig:finder}, we show the position of the IFU field on an archive \textit{HST} F656N image (GO 6423, P.I.\ D.\ Hunter; see Table~\ref{tbl:HST}) obtained with the Wide Field Planetary Camera 2 (WFPC2). Cluster 10 is located on the easternmost part of the ionized complex centred on H\two\ region 2 \citep{waller91} and is ideal for the study of the cluster--ISM interaction. Just off the south-west edge of the IFU field are two tentative 1.44MHz detections M-c and M-d of \citet{greve02}. To the north of the field, $\sim$25~pc from cluster 10, is VLA-16, a non-thermal radio source also reported by \citet{greve02} who consider it may be an extended low surface brightness supernova remnant (SNR).

%%%%%%%%%%%%%%%%%%%%%%%%%%%%%%%%
\subsection{Data Reduction}\label{sect:reduc}

Because of the complexity of the data products from IFU instruments, a considerable amount of time was invested in learning and developing reduction and analysis tools to enable us to extract the most information we could from this extensive dataset. 

The field-to-slit mapping for the GMOS IFU reformats the layout of the fibres on the sky to one long row containing five blocks of 100 object fibres interspersed by five blocks of 50 sky fibres. In this way, all 750 spectra can be recorded on the detector at the same time.

Basic reduction was performed using the Gemini pipeline reduction package (implemented in {\sc iraf}\footnote{The Image Reduction and Analysis Facility ({\sc iraf}) is distributed by the National Optical Astronomy Observatories which is operated by the Association of Universities for Research in Astronomy, Inc. under cooperative agreement with the National Science Foundation.}). There are a number of tasks within this package that are designed to perform certain functions, and each contain switches to allow them to be applied at different stages of the reduction procedure. The first stage was to run the {\sc gfreduce} task on the flat-fields, twilight flats and arc calibration exposures in order to prepare the raw files for reduction, and subtract the overscan level and bias frames. {\sc gfextract} traces and extracts the spectrum of each fibre (using a $\pm$2.5 pixel aperture in the spatial direction), and by running this on the flat-field image, produces a reliable trace of the position of each spectrum on the CCD. We then ran {\sc gfresponse} on the {\sc gfreduce}'d twilight flats to create a throughput correction function for each fibre (an essential step to correct differences in fibre-to-fibre transmission). The final step before reducing the science data was to extract the 750 arc calibration spectra from the arc lamp exposures (using {\sc gfreduce}) and establish and apply a wavelength calibration solution using {\sc gswavelength} and {\sc gftransform}.

Running {\sc gfreduce} on the science observations using the spectrum trace created from the flat-field, the throughput correction determined from the twilight flat, and the wavelength calibration from the arc exposures, creates a data file containing 750 reduced spectra, each one pixel in width and ordered by the position they were recorded on the CCD (which we will call the `fibre order'). After extraction of the spectra, the $x$ and $y$ spatial unit is now termed `spaxel' to differentiate from `pixel' which refers to the CCD.

Cosmic-rays were cleaned from the data at this stage with the Laplacian cosmic-ray identification routine {\sc lacosmic} \citep{vandokkum01}. Fits to the flux of the [O\one]$\lambda 5577$ telluric emission line, assumed to remain spatially stable in intensity for each observation set, for each of the 750 object spectra, showed a significant ($\sim$10 per cent) dip from the centre to the edges when plotted against fibre order. We decided that a second throughput correction was needed to remove this variation. This correction was measured by fitting a fourth order polynomial to the intensity curve, inverting the profile, normalising it to unity, then dividing this into the data using the {\sc iraf} task {\sc imarith}. 

Fits to the full-width half-maximum (FWHM) of the [O\one]$\lambda 5577$ line also showed unexpected variations with fibre order. We found that the groups of 50 sky fibres at either end of the CCD showed elevated widths compared to the central groups. Thus to be safe we extracted only sky fibres in the central blocks (i.e.\ fibres 200--250, 350--400 and 500--550). After extracting and averaging these fibres for each science frame, the resulting spectrum was subtracted from the rest of the spectra using {\sc imarith}. Because of diffuse emission from the galaxy extending past the 1 arcmin separation of the object and sky fields, there is the possibility that nebular emission could contaminate the sky spectra. To check the level of such contamination, we compared the flux of the H$\alpha$ line in representative spectra with and without sky subtraction. The difference in the flux for each case was 1--2 per cent and therefore deemed insignificant.

\subsection{Flux calibration} \label{sect:flux_calib}
As part of the calibration dataset, the flux standard star G191-B2B \citep{oke90} was observed using the same instrument set-up as described in Section~\ref{sect:reduc}. Reduction of these data followed the same procedure as described above for the science frames, except that we used the throughput response file and wavelength solution determined from the reduction of one of the other IFU fields. In order to obtain the optimum S/N spectrum for the standard star, we summed all object spaxels across the field of view using {\sc gfapsum} to create one averaged spectrum of the standard star. The sensitivity function was computed using the {\sc gsstandard} task (which includes a second-order extinction correction relevant to the Gemini-N observatory), and the flux calibration for each science frame was performed by the {\sc gscalibrate} task. 

Final combination of the individual exposures for each position was done using {\sc imcombine}. The resulting data file was then formed of 750 reduced, sky-subtracted and flux-calibrated spectra, one corresponding to each spaxel in the field-of-view (including the sky field). Separation of the sky fibres from the data files was achieved using the E3D Visualisation Tool \citep{sanchez04}, the development of which was a primary goal of the European Commission's Euro3D Research Training Network. Each science file now only contained the 500 object spectra.

In order to determine an accurate measurement of the instrumental contribution to the line broadening, we selected spectral lines from a wavelength calibrated arc exposure that were close to the H$\alpha$ and [O\three]$\lambda 5007$ lines in wavelength, and sufficiently isolated to avoid blends. We then fitted single Gaussians to these lines for all 750 apertures and took the average. The instrumental broadening (velocity resolution) of the final dataset varies between FWHM = 74\,$\pm$\,5~\kms\ at the blue end (with an average S/N of 4.6 in the continuum), to 59\,$\pm$\,2~\kms\ at the red end (with an average S/N of 14.5).

\subsection{Differential Atmospheric Refraction (DAR) Correction}
When observing the spectrum of an object, its light is refracted by the Earth's atmosphere by varying amounts at different wavelengths. This effect, termed differential atmospheric refraction (DAR), is therefore a function of wavelength (horizontal) and airmass (vertical). An advantage of IFS over traditional long-slit methods is that it is possible to determine and correct the effects of DAR using an \emph{a posteriori} procedure \citep[e.g.][]{arribas99}. In order to correct our data for DAR, we first had to convert each dataset to standard cube format. This involves changing the way in which the data are stored, from one spectrum (wavelength vs.\ flux) per spaxel (spatial coordinate) to a 2-D flux image at each wavelength point. For the DAR correction procedure, it is required that the data are sampled evenly in $x$ and $y$, and since the format of the GMOS fibres is hexagonal we had to apply an interpolation routine to resample the data into contiguous squares with an equivalent `diameter'. A diameter of $0\farcs18$ gave the closest match to the \emph{number} of original spaxels when interpolating from hexagons to squares.

Measurement of the DAR shift can be achieved easily and automatically if there is an unresolved point source in the field-of-view: in our case cluster 10. With the data in cube format (2 spatial axes and one of wavelength) and using scripts provided by L.\ Christensen (ESO), the spatial shifts were traced by cutting the cubes into a series of monochromatic images, and fitting a two-dimensional Gaussian function to the unresolved star cluster to determine the centroid position. A third order polynomial was used to fit this trace (with wavelength) through the datacubes, giving the relative shifts in RA and Dec for each slice with respect to the first (reference) wavelength. The measured offsets were then used to shift each wavelength slice by then required amount, and were the combined back into a datacube with the original grid sampling using a modified version of the {\sc iraf} {\sc stsdas}\footnote{{\sc stsdas} is the Space Telescope Science Data Analysis System; its tasks are complementary to those in {\sc iraf}.} {\sc drizzle} task \citep[originally written as an implementation of the image combination method known as variable-pixel linear reconstruction;][]{fruchter02}. The result is a datacube where every flux at each point in the object corresponds to the same position in the image for each wavelength point.

The final stage was to crop the edges of each field in order to remove spaxels only containing flux for part of the wavelength range. This is an inevitable side-effect of performing a DAR correction.

\subsection{HST images}

\begin{table*}
\centering
\caption {HST archive images of NGC 1569.}
\label{tbl:HST}
\begin{tabular}{l l l l l l}
\hline
Filter & Camera/CCD & Plate Scale & Exposure & Programme & P.I. \\
& & ($''$ pix$^{-1}$) & Time (s) & I.D. \\
\hline
F330W & ACS/HRC & 0.026 & $220\times 3$ & 9300 & H.\ Ford \\
F439W & WFPC2/PC & 0.045 & $700\times 16$ & 6111 & C.\ Leitherer \\
F555W & ACS/HRC & 0.026 & $130\times 3$ & 9300 & H.\ Ford \\
F656N & WFPC2/PC & 0.045 & $800\times 2$ & 6423 & D.\ Hunter \\
F814W & ACS/HRC & 0.026 & $130\times 3$ & 9300 & H.\ Ford \\
\hline
\end{tabular}
\end{table*}

To supplement our studies, we obtained {\it HST} ACS/HRC and WFPC2 broad- and narrow-band images of the central regions of NGC 1569 from the {\it HST} archive. These images are listed in Table~\ref{tbl:HST} (together with the corresponding program IDs and P.I.) and were calibrated using the ``on-the-fly'' pipeline system. Dithered images were first registered using their WCS (World Coordinate System) information, then combined using the {\sc iraf} {\sc imcombine} routine in order to remove cosmic-rays.

\begin{figure*}
\begin{minipage}{4.5cm}
\vspace*{-0.6cm}
\includegraphics[width=4.5cm]{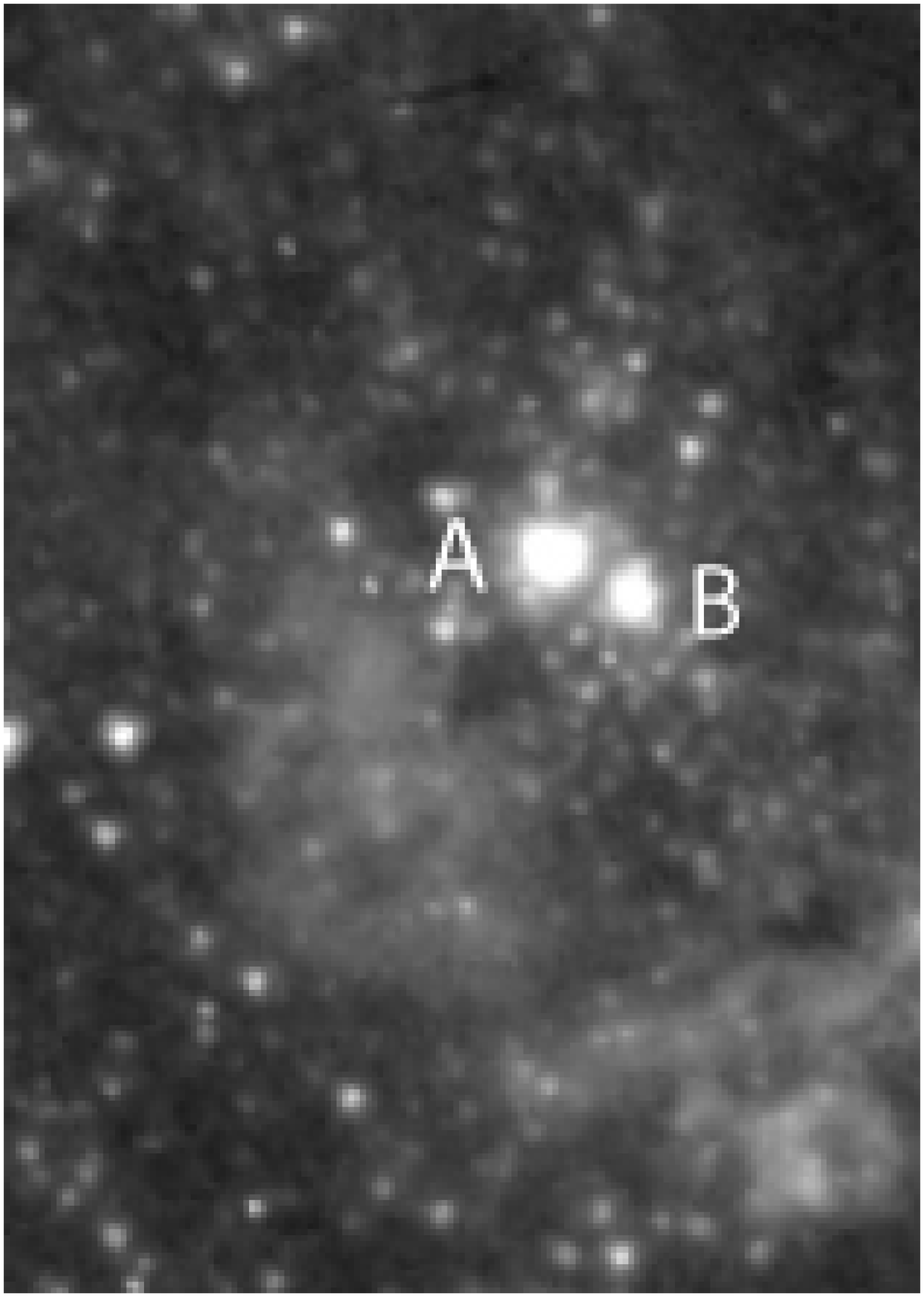}
\end{minipage}
\hspace*{0.3cm}
\begin{minipage}{5.5cm}
\includegraphics[width=5.5cm]{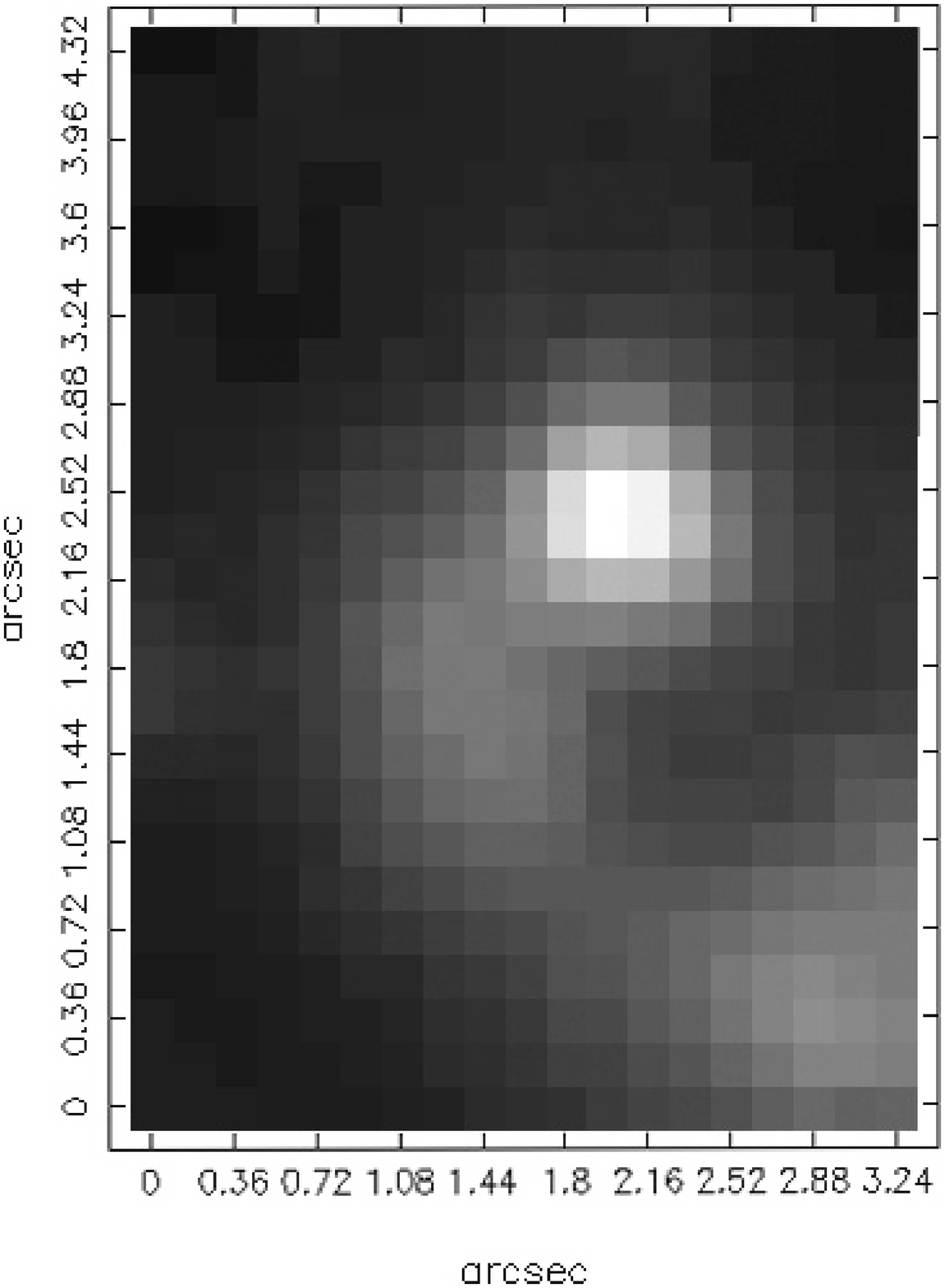}
\end{minipage}
\hspace*{0.3cm}
\begin{minipage}{5.5cm}
\includegraphics[width=5.5cm]{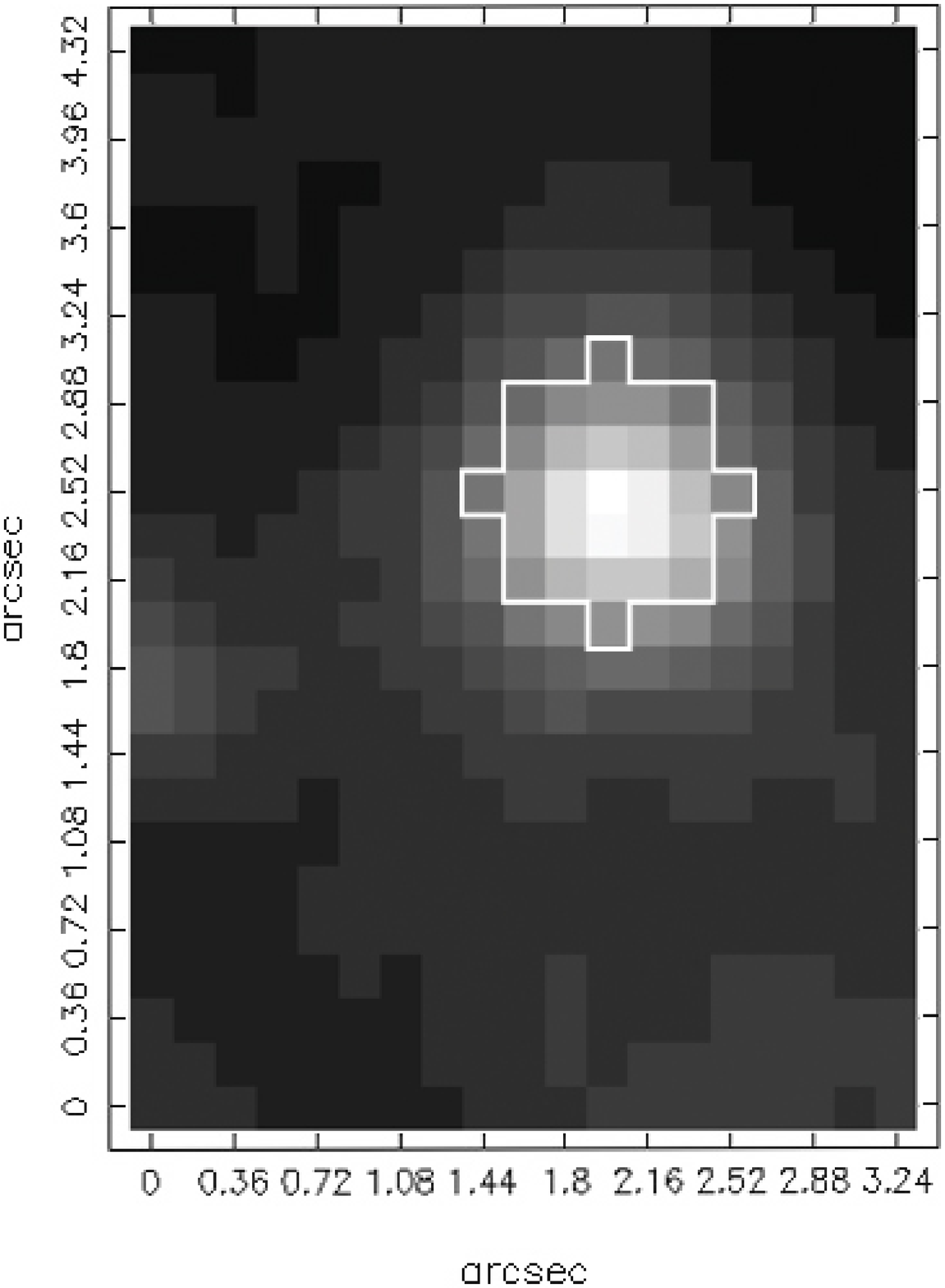}
\end{minipage}
\caption{\emph{Left:} \textit{HST}/ACS HRC F555W image of the region covered by the IFU pointing. The two components of cluster 10 are labelled. \emph{Centre:} IFU data convolved with a $V$-band filter function. \emph{Right:} flux map in the continuum only (6645--6660~\AA), showing the position of cluster 10 and the spaxels extracted to form the cluster spectrum. North is up and east is left.}
\label{fig:acs&cont}
\end{figure*}

Fig.~\ref{fig:acs&cont} introduces the concept of imaging with IFU data. On the left is the {\it HST} ACS/HRC F555W image of the region covered in our IFU pointing. In the centre we show a `virtual' $V$-band image reconstructed from our IFU data by convolving the spectrum from each spaxel with a $V$-band filter function. The similarity between the two images is obvious once the resolution of each is taken into account (ACS/HRC = $0\farcs026$ pixel$^{-1}$; GMOS IFU = $0\farcs18$ spaxel$^{-1}$ but seeing-limited at $\sim$$0\farcs8$). One must keep in mind the fact that the ACS F555W filter contains contamination from nebular [O\three] emission, which is why a certain amount of nebulosity is seen surrounding cluster 10. The right-hand plot is of the same field but restricted in wavelength to cover the continuum only (in the range 6645--6660~\AA). Overlaid is an outline of the 24 spaxels summed to create a spectrum of cluster 10 (Section~\ref{sect:clus10}). 

%%%%%%%%%%%%%%%%%%%%%%%%%%%%%%%%
\section{Cluster 10} \label{sect:clus10}

Cluster 10 is the third visually brightest cluster identified by \citet{hunter00} but, as can be seen from Fig.~\ref{fig:acs&cont} (left panel), it is in fact a double cluster. Because of the close proximity of the two sources, previous studies \citep[e.g.][]{hunter00} have treated cluster 10 as one object. Fitting the light profiles across the length and breadth of the IFU field with a Gaussian profile yields a average FWHM of 4.7 spaxels or $0\farcs84$. This corresponds to the maximum seeing disc size at the time of observation, meaning that the two sub-clusters are not resolved in our IFU data.

From the F555W ACS/HRC image, we measure the separation of the two cluster 10 components as $0\farcs35$ (3.7~pc). The coordinates of the two sources are given in Table~\ref{tbl:clus10}. Measurement of the size of the two clusters was made on the ACS/HRC F555W image, and was achieved using {\sc ishape} \citep{larsen99b} together with the {\sc tinytim} package \citep{krist04} to correct for the point-spread function (PSF). Using a circular Moffat function with a power index of 1.5, and a fit radius of $0\farcs 2$ gives an effective radius, $R_{\rm eff} = 1.18\pm 0.05$~pc for the north-eastern component (hereafter cluster 10A; see Fig.~\ref{fig:acs&cont} left panel), and $R_{\rm eff} = 0.88\pm 0.15$~pc for the south-western component (hereafter cluster 10B). Due to this very small separation, and the crowding of the field around cluster 10 with other fainter sources, standard aperture photometry is not possible (and we cannot measure absolute magnitudes). However if all that are needed is relative magnitudes, then small-radius aperture photometry is sufficient. Using a radius of 4 pixels for the HRC images (F333W, F555W and F814W) and 2.5 pixels for the WFPC2 image (F439W) and suitable background annuli, we measure (F555W--F814W) = 1.12 and (F330W--F439W) = $-0.27$ for the cluster 10A, and (F555W--F814W) = 0.42 and (F330W--F439W) = $-0.90$ for cluster 10B. We note that since the cluster radii are so similar, using the same aperture for both does not introduce significant errors.

\begin{figure}
\includegraphics[width=0.48\textwidth]{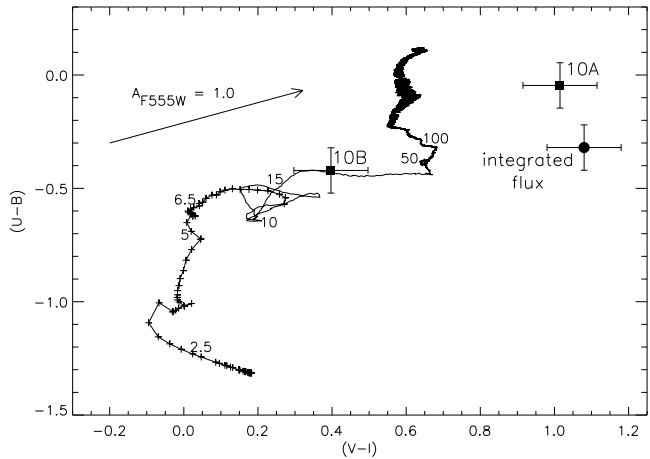}
\caption{Colour-colour diagram showing (solid line) \textsc{Starburst99} instantaneous burst models for $Z=0.004$. Crosses mark age points on the evolutionary curve from 5--7~Myr; some significant age steps are labelled (units of Myr). The arrow represents the extinction vector for $A_{\rm F555W}=1.0$. Solid squares show the colours of the two clusters (cluster 10A and 10B); the solid circle shows the colour of the integrated flux of the two.}
\label{fig:clus_colour}
\end{figure}

Fig.~\ref{fig:clus_colour} shows the evolutionary path of a $Z=0.004$ \textsc{Starburst99} \citep[\textsc{sb99};][]{leitherer99} instantaneous burst evolutionary-synthesis model with a standard Kroupa initial mass function (IMF) in $U-B$, $V-I$ colour space, together with the colours that have been determined for both cluster 10 components. Also plotted is the colour of the integrated flux of the two sources \citep[measured using an aperture equivalent in to the cluster size found by][]{hunter00} to compare to parameters derived from the spectrum, since the spectrum is derived from a convolution of the light from both components. \textsc{SB99} does not give outputs in terms of \textit{HST} filters, so the observed photometry was converted from \textit{HST} to Johnson magnitudes using the conversion factors given by \citet{holtzman95} for the WFPC2 data, and \citet{sirianni05} for the ACS data. From the diagram we can now derive an approximate age and an estimate of the reddening for each of the cluster components simultaneously. Assuming a extinction value sufficient to deredden each point until it intercepts the model track has the following results. For cluster 10A, an $A_{V}$ of $\approx$0.8 [$E(B-V)\approx 0.26$] corresponds to an age of 50--100~Myr, whereas $A_{V}\approx 1.2$--2 [$E(B-V)\approx 0.65$--0.40] gives an age of 5--20~Myr. For cluster 10B, zero extinction would imply an age of $\sim$20~Myr, an $A_{V}$ of $\approx$0.3 [$E(B-V)\approx 0.10$] intercepts the track at $\sim$7--15~Myr, whereas an $A_{V}\approx 0.8$ [$E(B-V)\approx 0.25$] implies an age of $\sim$5~Myr. The extinction required for the combined point to cross the track at $\sim$5--7~Myr is $A_{V}\approx 1.6$ [$E(B-V)\approx 0.54$].

Adopting a Galactic foreground reddening of $A_{V} = 1.64$ [$E(B-V) = 0.53$] \citep[following][]{origlia01, relano06}, we can immediately rule out ages requiring reddening values less than this. We therefore conclude that 10A must have an age of 5--7~Myr (with a reddening in addition to the Galactic foreground level of $A_{V}\approx 0.0$--0.4), and 10B an age of $\leq$ 5~Myr (consistent with zero reddening after foreground correction). For the colour of the combined point to match these age determinations, the reddening value required is consistent with purely Galactic effects.

A further test to obtain an estimate of the cluster age is to compare the equivalent width of the H$\alpha$ emission line to the predictions of evolutionary synthesis models. Since we only have a spectrum of the combined light from both the clusters, we can only put an upper limit on the flux-weighted age. Measuring an equivalent width of H$\alpha=91\pm 3$~\AA\ and using the same \textsc{sb99} model as used above ($Z=0.004$, Kroupa IMF), this gives an upper limit to the age of $\sim$6.5~Myr, and is consistent with the values derived above.

Now we have an estimated age, we can derive an estimate for the mass by comparing the absolute magnitude, $M_{\rm F555W}$, to evolutionary synthesis models. As mentioned above, we cannot measure an absolute magnitude of the individual sub-clusters, but using an aperture large enough to include both, we find a combined $M_{\rm F555W, cluster 10} = -9.61$~mag. Converting to Johnson magnitudes, we find $M_{V, {\rm cluster 10}} \approx -9.2$~mag, and comparing our measurements to a \textsc{sb99} 6~Myr, $1\times 10^{6}$~\Msol{} model results in a predicted mass of 2--$5 \times 10^{3}$~\Msol{} (depending on the IMF formulation used) for the combined mass of both sub-clusters. Since there are uncertainties involved in transforming from the \textit{HST} to Johnson magnitude system, we compared our measured $M_{\rm F555W}$ to the absolute magnitudes predicted by the equivalent \citet{bc03} model (for which results in \textit{HST} filters are given). The derived mass is 7--$13\times 10^{3}$~\Msol{} (depending on IMF), which is in general agreement with the \textsc{sb99} result. We therefore adopt the value of $7\pm5 \times 10^{3}$~\Msun\ for the combined mass of the two subclusters, meaning that neither of them can be defined as a super star cluster in the generally accepted definition.

\citet{buckalew00} conducted a study of NGC 1569 using \textit{HST} WFPC2 narrow-band imaging of the He\two\,$\lambda$4686 line. They discovered fifteen He\two{} emitting sources across the galaxy, of which five were associated with stellar clusters including cluster 10 (specifically only the south-western component; 10B). They attribute the emission as nebular in origin \citep[after][who found wide-spread nebular He\two{} emission in this region]{kobulnicky97}. However He\two{} is also strongly emitted in the atmospheres of Wolf-Rayet (WR) stars of both WC and WN type. \citet{buckalew00} comment that \emph{if} the emission they detect in cluster 10 was attributed fully to WR stars, it would correspond to the equivalent of three WNL-type stars \citep[using the calibration of][]{vacca92}. WR stars exist in clusters between ages of approximately 3--5~Myr, so in light of our age estimates for the two cluster components derived above, a WR origin for the He\two{} emission in cluster 10B is perfectly reasonable.

\subsection{Cluster Spectrum} \label{sect:clus_spec}
\begin{figure*}
\includegraphics{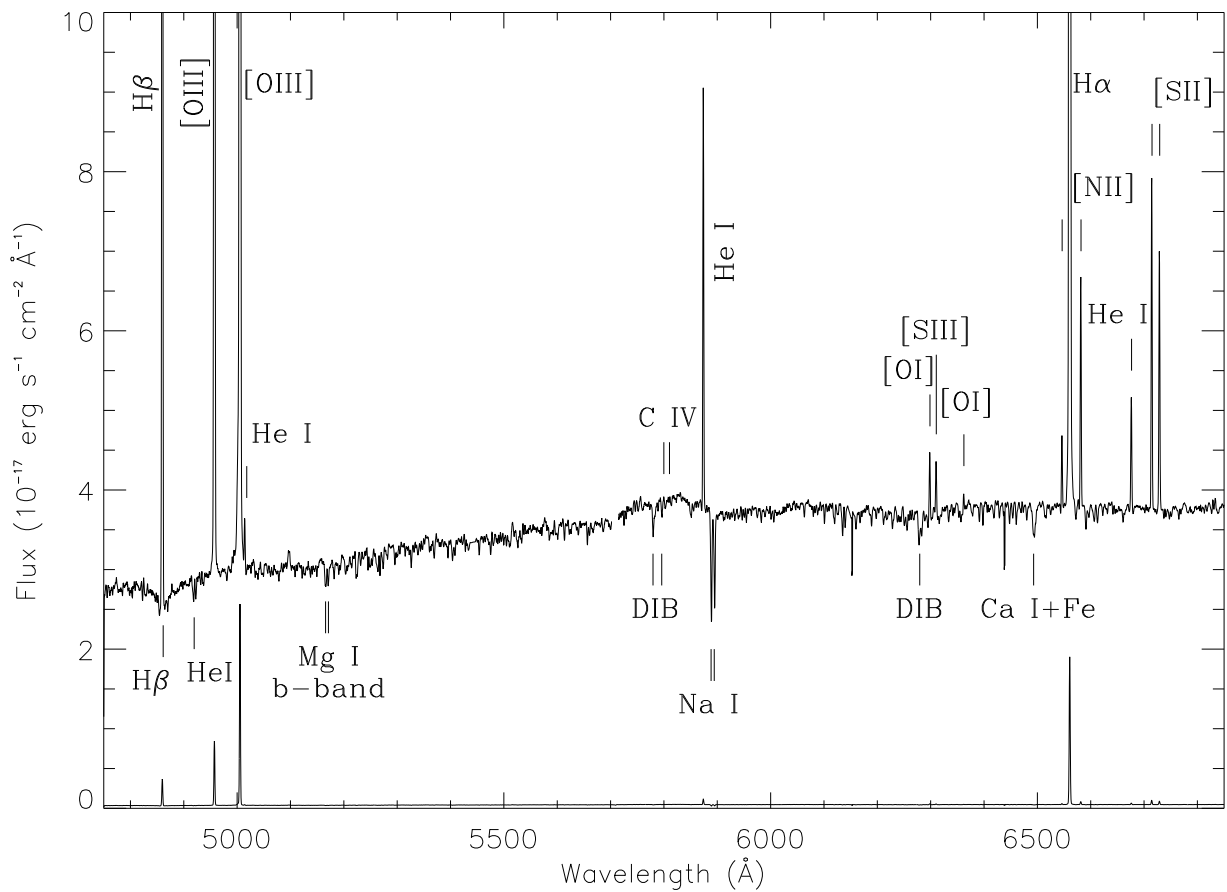}
\caption{Plot of the extracted spectrum of cluster 10 (smoothed by 5~\AA). All identified absorption and emission features are labelled. The lower line corresponds to a $\times$80 reduction of the main plot to illustrate the the spectrum with a full flux range. Notice the strength of the [O\three]$\lambda 5007$ emission compared to the hydrogen recombination lines.}
\label{fig:clus_spec}
\end{figure*}

To extract the spectrum of cluster 10, we summed 29 spaxels (covering an area with an equivalent diameter to the FWHM of the cluster measured from the IFU continuum image) centred on the continuum source as shown on the right-hand plot of Fig.~\ref{fig:acs&cont}. The resulting spectrum is shown in Fig.~\ref{fig:clus_spec}.

The spectrum is dominated by strong emission lines arising from the nebula surrounding the cluster, but also contains a broad emission feature that we identify with C\four$\,\lambda\lambda$5801,5812 arising from WC-type Wolf-Rayet stars. We also detect a large number of weak stellar absorption lines, diffuse interstellar bands (DIBs) and interstellar absorption lines. The detection of a red WR-bump is not unsurprising considering the age we find for cluster 10B, and strengthens the argument for a WR origin for the He\two{} emission detected by \citet{buckalew00}. Further evidence of the young nature of cluster 10 comes from the high ratio of [O\three] to H$\beta$ (see Fig.~\ref{fig:clus_spec}, lower line), and the presence of strong [S\three] emission, indicating a high level of ionization in the surrounding H\two{} region, which can only arise in the presence of young O stars.

Simultaneously fitting the WR feature with the superimposed absorption lines, we measure the bump to have a FWHM of $\sim$130~\AA, a total luminosity of $2.5\times 10^{37}$~erg~s$^{-1}$ (assuming a distance of 2.2~Mpc and $A_{V} = 1.64$), and an equivalent width of $12\pm 2$~\AA. This measured FWHM is unusually broad compared to canonical measurements \citep*{crowther98}, and is likely to result from inaccuracies in the fit due to contamination from the superposed DIB absorption lines. \citet{crowther06} present measurements of WR bump fluxes for a number of WC4 type stars in the LMC (the closest galaxy containing a significant population of WR stars at low metallicity -- $Z=0.008$) which can be used to determine the equivalent number of stars in our observation. They find one LMC WC4 star to have a red WR-bump luminosity of 3.3$\pm$$1.6\times 10^{36}$~erg~s$^{-1}$, thus the luminosity equivalent number of WC4 stars in our observed bump is estimated to be $10\pm 5$, once all the uncertainties have been taken into account. For comparison, \citet{g-d97} find a luminosity equivalent of $20\pm 10$ WNL stars in SSC A by measuring the flux in the blue WR-bump.

By fitting Gaussian profiles to all the other emission and absorption lines present in the spectrum, we measured the central velocity and FWHM of each, and present them in Table~\ref{tbl:line_id}. The average radial velocity of the nebular emission lines is $-82\pm 4$~\kms{}. Since our cluster spectrum only contains blended photospheric absorption lines, it is not possible to measure an accurate velocity for the stellar component. The evolutionary synthesis models of \citet{bc03} provide a set of spectral energy distributions (SEDs) calculated for similar input ranges as the \textsc{sb99} models, and although being less accurate at predicting very young ages, their SEDs are derived from actual observed spectral libraries rather than model stellar atmospheres. Their inclusion of observed stellar spectra at high resolution therefore make them ideal for comparing with our spectrum to derive a cluster velocity. Using a 0.25~\Zsol{}, 5~Myr instantaneous burst model in the wavelength range 5100--5600~\AA{} (where many faint stellar absorption lines are present), we find a match at $-85\pm 5$~\kms{}. A good correlation at this velocity is also found at the position of the Ca\one+Fe blend at 6494~\AA, which is thought to be photospheric in origin \citep{heckman95}. As well as stellar absorption lines, we detect the interstellar (IS) absorption lines of Na\one\,$\lambda \lambda$5890,5896 and a number of DIBs \citep{herbig95, heckman00b}. Their velocities are given in Table~\ref{tbl:line_id}, and are consistent with being Galactic in origin (average velocity $-40\pm 5$~\kms), although the Na\one{} profiles also show blue wings presumably associated with IS absorption within NGC 1569.

Determining the proper heliocentric systemic velocity, $v_{\rm sys}$, of NGC 1569 has proven to be a difficult task due to its irregular and disturbed morphology. The first study of \mbox{NGC 1569's} H\one\ gas distribution by \citet{reakes80} concluded a $v_{\rm sys}=-77$~\kms. CO observations by \citet{taylor99} found the same value, whilst \citet*{tomita94} and \citet{heckman95} found an H$\alpha$ $v_{\rm sys}$ of $-90$~\kms\ and $-68$~\kms\ respectively from their long-slit data. A more recent, high resolution H\one{} study of the NGC 1569 system was made by \citet{muhle05}, who found the peak of the H\one\ profile at the position of cluster 10 to have a velocity of $-80$~\kms{} (S.\ M\"uhle, private communication). Thus, since our measurements of the ionized gas and cluster velocities are also consistent with this value, we adopt a systemic velocity for the cluster 10 region of $v_{\rm sys, 10}=-80$~\kms.

\renewcommand{\baselinestretch}{1.2}
\begin{table}
\begin{center}
\caption{Absorption and emission line measurements for cluster 10 (all lines are emission unless otherwise stated). Widths are corrected for instrumental contribution and velocities are quoted in the heliocentric frame of reference. Typical errors on the velocities of the emission lines are $\sim$5--10~\kms, and on the widths $\sim$0.5--5~\kms{}. Errors for the absorption lines are listed individually.}
\label{tbl:line_id}
\begin{tabular}{l l l r @{$\;$} l }
\\
\hline
Line & \multicolumn{1}{c}{$\lambda_{\rm air}$} & Velocity & \multicolumn{2}{c}{FWHM} \\
& \multicolumn{1}{c}{(\AA)} & (\kms) & \multicolumn{2}{c}{(\kms)} \\
\hline 
H$\beta$ & 4861.33 & $-84.4$ & 50.7 \\
H$\beta$ abs\,$^{a}$ & 4861.33 & $-55\phantom{.0}\pm 40$ \\
He\one{} abs\,$^{a}$ & 4921.93 & $-84\phantom{.0}\pm 8$ \\
{[}O\three{]} & 4958.92 & $-79.0$ & 44.7 \\
{[}O\three{]}\,$^{b}$ & 5006.84 & $-75.1$ & 31.6 \\
He\one\,$^{c}$ & 5015.68 & $-74.7$ & 42.3 \\
Mg\one{} abs & 5167.32 & $-77\phantom{.0}\pm 7$\\ %& 180 & $\pm\; 15$ \\
Mg\one{} abs & 5172.68 & $-89\phantom{.0}\pm 6$\\ %& 100 & $\pm\; 15$ \\
DIB & 5780.5 & $-40.9\pm 10$ & 160 & $\pm\; 35$ \\
DIB & 5797.0 & $-47.5\pm 20$ & 70 & $\pm\; 25$ \\
He\one & 5875.67 & $-82.7$ & 33.9 \\
Na\one{} abs & 5889.95 & $-38\phantom{.0}\pm 8$ \\
Na\one{} abs & 5895.92 & $-34\phantom{.0}\pm 8$ \\
DIB & 6283.9 & $-41.0\pm 10$ & 110 & $\pm\; 20$ \\
{[}O\one{]} & 6300.30 & $-84.9$ & 32.8 \\
{[}S\three{]} & 6312.10 & $-82.0$ & 50.5 \\
{[}O\one{]}$^{a}$ & 6363.78 & $-80.1$ \\
{[}N\two{]} & 6548.03 & $-81.4$ & 28.3 \\
H$\alpha$$^{b}$ & 6562.82 & $-89.1$ & 38.8 \\
{[}N\two{]} & 6583.41 & $-82.8$ & 44.5 \\
He\one & 6678.15 & $-81.6$ & 44.1 \\
{[}S\two{]} & 6716.47 & $-85.1$ & 40.8 \\
{[}S\two{]} & 6730.85 & $-84.4$ & 46.5 \\
\hline
\end{tabular}

\begin{tabular}{p{6cm}}

$^{a}$ Line too weak or confused to measure an accurate FWHM \\
$^{b}$ Narrow component (C1) only \\
$^{c}$ Blended with [O\three]
\end{tabular}
\end{center}
\end{table}
\renewcommand{\baselinestretch}{1.5}

A check on the value of reddening for the combined light of both cluster components can be made by comparing the observed H$\alpha$/H$\beta$ flux ratio with the intrinsic case B values from \citet{humstor87}. We derive an absorption-corrected reddening of $E(B-V)=0.54\pm 0.15$~mag from Gaussian fits to the H$\alpha$ emission and H$\beta$ emission and absorption lines. This value is consistent with both the adopted value of Galactic foreground reddening and with the extinctions derived from the colour-colour plot analysis given above. We therefore conclude that cluster 10 only suffers from Galactic foreground reddening, i.e.\ very little of its light is attenuated by material in NGC 1569 itself. The flux ratio of single Gaussian fits to the [S\two]$\lambda \lambda$6717,6731 emission lines gives a value of $1.25^{+0.02}_{-0.01}$ hence an electron density, $n_{\rm e} = 175^{+18}_{-20}$~cm$^{-3}$ (assuming an electron temperature, $T_{\rm e} = 10^{4}$~K). These results are summarised in Table~\ref{tbl:clus10}.

\renewcommand{\baselinestretch}{1.2}
\begin{table}
\begin{center}
\caption{Observed and derived properties for cluster 10.}
\label{tbl:clus10}
\begin{tabular}{l c c}
\\
\hline
& \multicolumn{2}{c}{Cluster component} \\
Parameter & 10A & 10B \\
\hline
Adopted systemic velocity & \multicolumn{2}{c}{$-80$} \\
\hspace{0.5cm}(\kms) \\
Distance (Mpc) & \multicolumn{2}{c}{$2.2\pm 0.6$\,$^{\dag}$} \\
Coordinates (J2000) & $04^{\rm h}\,30^{\rm m}\,47\fsec16$ & $04^{\rm h}\,30^{\rm m}\,47\fsec12$ \\
& $+64^\circ\,51'\,00\farcs8$ & $+64^\circ\,51'\,00\farcs6$ \\
(F555W--F814W) & 1.12 & 0.42 \\
(F330W--F439W) & $-0.27$ & $-0.90$ \\
Half-light radius, R$_{\rm eff}$ (pc) & $0.88\pm 0.05$ & $0.60\pm 0.15$ \\
Total $E(B-V)$ & \multicolumn{2}{c}{$0.55\pm 0.15$} \\
Electron density, $n_{\rm e}$ (cm$^{-3}$) & \multicolumn{2}{c}{175$^{+18}_{-20}$} \\
Age (Myr) & 5--7 & $\leq$\,5 \\
$M_{\rm F555W}$ (mag) & \multicolumn{2}{c}{$-9.61$} \\
Mass (\Msol) & \multicolumn{2}{c}{$7\pm 5 \times 10^{3}\,^{\ddag}$} \\
%No.\ of equiv.\ O7V stars & \multicolumn{2}{c}{$30\pm 20$} \\
No.\ of equiv.\ WC4 stars & \multicolumn{2}{c}{$10\pm 5$} \\
Radial velocity (\kms) & \multicolumn{2}{c}{$-85\pm 5$} \\
\hline
\end{tabular}

\begin{tabular}{p{7.8cm}}

$^{\dag}$ \citet{israel88} \\
$^{\ddag}$ Depending on whether a Salpeter or Kroupa IMF is used
\end{tabular}
\end{center}
\end{table}
\renewcommand{\baselinestretch}{1.5}

%%%%%%%%%%%%%%%%%%%%%%%%%%%%%%%%
\section{Properties of the Ionized Gas}\label{sect:profile}

We now discuss the IFU data relating to the ionized gas in the environment of cluster 10. The signal-to-noise and spectral resolution of these data are sufficiently high to resolve multiple Gaussian components to each emission line in each of the 500 spectra across the field-of-view. Before discussing our findings, we first describe the methods used to fit the spectral line profiles and visualise the results.

\subsection{Decomposing the line profiles}\label{sect:pan}
The automated fitting of multiple profiles to 500 spectra each containing eight emission lines has proved challenging. For this task we have adopted an IDL-based, general-purpose curve-fitting package, called \textsc{pan} \citep[Peak ANalysis;][]{dimeo}. It was written with the aim of allowing the user to visually interact with the program through the use of IDL's sophisticated `widget-based' toolkits, and uses the Levenberg-Marquardt technique to solve the least-squares minimisation problem. The primary features that caused us to favour this program over any other was its ability to read in multiple spectra at once in an array format, and its method of determining the initial guess parameters of a model profile (namely allowing the user to visually and interactively specify amplitude, position and width). Originally written to analyse data from neutron-scattering experiments, the program has required some modification to suit our needs. We have had to add modules to convert our FITS format data into a format that \textsc{pan} can read; we have also had to modify the way it outputs the fit results and $\chi^{2}$ values to something that can easily be read by a plotting package.

To obtain good, reliable fits, the spectra read into \textsc{pan} had to contain accurate associated error arrays. These error arrays were calculated by taking the square-root of the raw counts in the pre flux-calibrated and throughput-corrected data, converting this to a percentage of the total raw counts, and applying this to the final reduced and calibrated data in order to obtain the uncertainty equivalent to the Poissonian noise. This step was included when converting from FITS to \textsc{pan} input format. By comparing our error estimates on fits to telluric features (see below) to simple Poissonian statistics, we find this to be a realistic measure of the cumulative uncertainties.

\begin{table}
\centering
\caption {Wavelength limits used in fitting each line}
\label{tbl:fitlim}
\begin{tabular}{c r @{\,--\,} l}
\hline
Spectral Line & \multicolumn{2}{c}{Wavelength (\AA)} \\
\hline 
H$\beta$ & 4840 & 4880 \\
{[}O\three{]}$\lambda5007$ & 4940 & 5020 \\
{[}O\one{]}$\lambda6300$ & 6250 & 6308 \\
{[}N\two{]}$\lambda6583$ & 6570 & 6595 \\
H$\alpha$ & 6550 & 6575 \\
{[}S\two{]}$\lambda6717$ & 6700 & 6724 \\
{[}S\two{]}$\lambda6731$ & 6720 & 6750 \\
\hline
\end{tabular}
\end{table}

Each line in each of the 500 spectra was fitted using a single, double and triple Gaussian component initial guess, where line fluxes were constrained to be positive and widths to be greater than the instrumental contribution (measured from wavelength calibrated arc exposures to be 74\,$\pm$\,5~\kms\ at [O\three] and 59\,$\pm$\,2~\kms at H$\alpha$; Section~\ref{sect:flux_calib}). The wavelength limits used in fitting each line are given in Table~\ref{tbl:fitlim}. In general, the shape of the emission line profiles is a convolution of a bright, narrow component overlying a faint, broad component (most obvious in high S/N lines, {e.g.~H$\alpha$). This is consistent with what \citet{heckman95} found for the central regions. For the double component fits, the initial guess was always made with the first Gaussian as the narrow component (hereafter referred to as C1), and the second Gaussian as the broader component (hereafter C2). For the triple component fit, the additional component (C3) was specified with a initial guess assigning it to a supplemental narrow line at the same wavelength as the main narrow line. This consistent approach helped limit the confusion that might arise during analysis as to which Gaussian fit belonged to which component of the line, as well as aiding the $\chi^{2}$ minimisation process.

\begin{figure*}
\centering
\includegraphics[width=\textwidth]{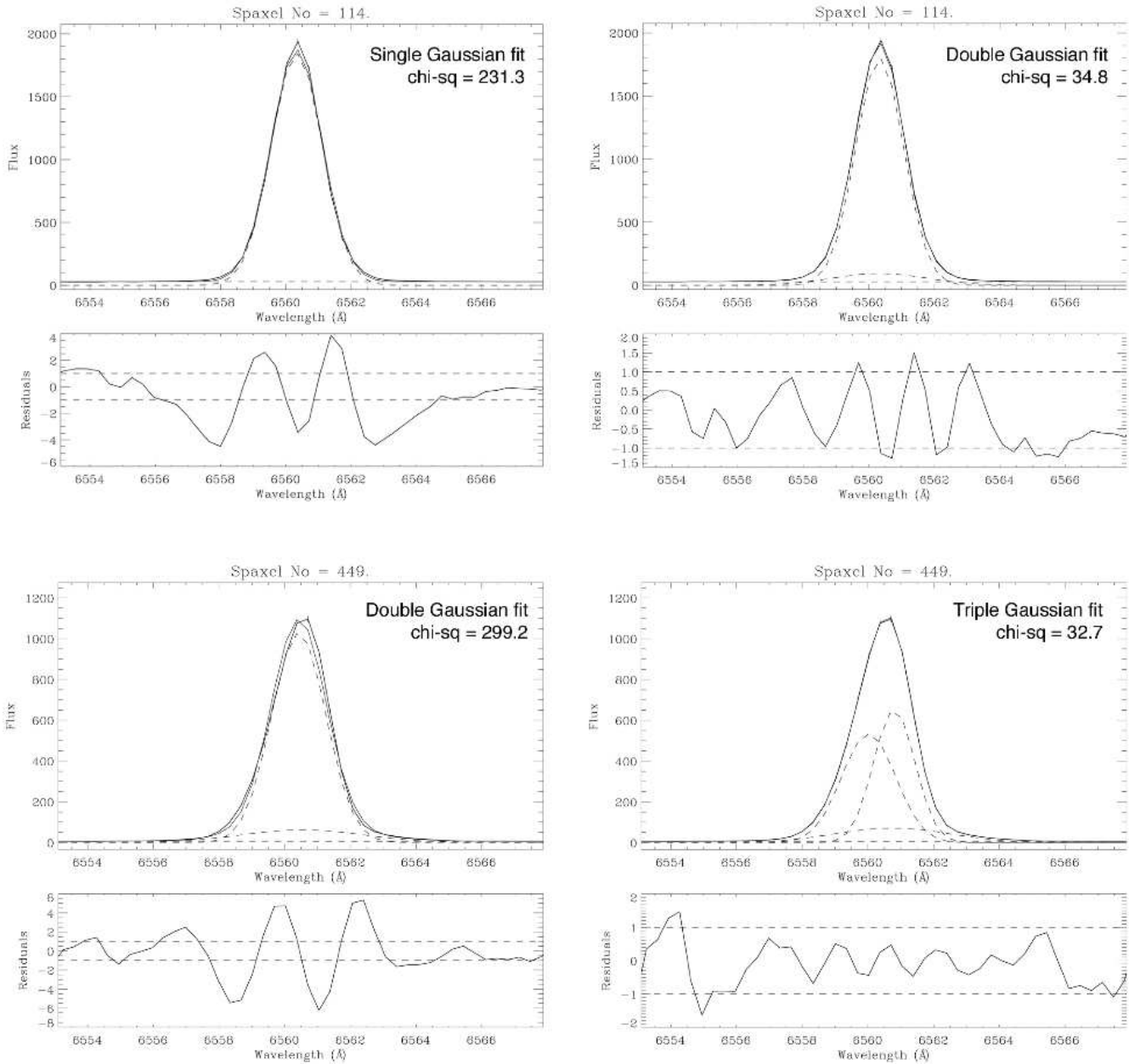}
\caption{Examples of H$\alpha$ line fits made by \textsc{pan} comparing a single and double Gaussian fit for spaxel 114 and a double and triple fit for spaxel 449 ($y$-axis in arbitrary flux units, but note the difference in scale). The black solid line is the observed data and the dashed lines represent the individual Gaussian components. Below each graph is a plot of the residuals (in units of $\sigma$) with dashed guidelines plotted at $\pm 1$ (see text for an explanation of how the residuals are calculated). By comparing the ratios of the $\chi^{2}$ values, the significance of a higher number of fit components can be tested.}
\label{fig:fitegs}
\end{figure*}

We used the statistical F-test to determine how many Gaussian components best fit an observed profile. The F-distribution is a ratio of two $\chi^{2}$ distributions, denoted by the degrees of freedom for the numerator $\chi^{2}$ and the denominator $\chi^{2}$. For statistically comparing the quality of two fits, this function allows one to calculate the significance of a variance ($\chi^{2}$) increase that is associated with a given confidence level, for a given number of degrees of freedom. The test will output the minimal increase of the $\chi^{2}$ ratio that would be required at the given confidence limit for deciding that the two fits are different. If the $\chi^{2}$ ratio is higher than this critical value, the fits are considered statistically distinguishable. The output files for the single, double and triple Gaussian models were passed through a script which applied the F-test to extract the results for the most suitable case for each spectral line. Fig.~\ref{fig:fitegs} shows two example H$\alpha$ line profiles, each from a different spaxel, with their corresponding Gaussian fits and residual plots shown. \textsc{pan} calculates the residuals, $r_{\rm i}$, using the following formula
\begin{equation}
r_{\rm i} = \frac{y_{\rm i}^{\rm fit} - y_{\rm i}^{\rm data}}{\sigma_{\rm i}}
\end{equation}
where $\sigma_{\rm i}$ are the uncertainties on $y_{\rm i}^{\rm data}$. The upper example represents an instance where two Gaussian components are needed to fit the profile---with the inclusion of a second Gaussian the fit $\chi^{2}$ is dramatically reduced and the residuals fall within the $\pm 1$ guidelines. The lower plots show an example of a profile which is best fit with a triple Gaussian model. The addition of a second narrow, bright component as well as the broad, faint component causes a significant reduction in the $\chi^{2}$ and again the residuals fall within the $\pm 1$ guidelines.

We also applied a number of physical tests to further improve the accuracy of the results. Firstly, for a fit to be accepted, the measured FWHM had to be greater than the associated error on the FWHM result (a common symptom of a bad fit), and secondly the FWHM had to be smaller than 15~\AA\ (to guard against fitting continuum or spurious features). We also applied some additional filtering in order to ensure the first Gaussian component always corresponds to our definition of the first component (i.e. if flux(C1) $<$ flux(C3) AND fwhm(C1) $<$ fwhm(C3) then swap C3 with C1). This only applies when the line contains three components and is sufficiently well resolved to reliably fit them all.

The bright [O\three]$\lambda$5007 line also exhibits the general line profile shape described above (narrow bright component with underlying fainter broad component), but unfortunately due to an instrumental artefact present at low levels in the blue wing of the line, we have not been able to characterise the shape of the secondary components accurately. The flux, width and velocity maps of the main bright [O\three]$\lambda$5007 component (C1) look very similar to the equivalent H$\alpha$ maps, so we do not show them here. Measurements of the [O\three]$\lambda$4959 line also result in similar flux, FWHM and velocity distributions, but since it is a weaker line we cannot fit the low intensity broad component with any confidence.

\subsection{How well can we fit each component?} \label{sect:errors}
The $\chi^{2}$ minimisation algorithm used by \textsc{pan} is very sensitive to the uncertainties associated with the input spectra. Because of this, we paid special attention to creating accurate Poissonian error arrays from the raw un-calibrated spectra, that we then associated with the fully reduced spectra for the purposes of fitting (see Section~\ref{sect:pan}). We made a number of tests comparing the results of Gaussian fits to high S/N arc lines made with \textsc{pan} with those output by other line fitting routines (such as {\sc elf} within the STARLINK {\sc dipso} program). We found that despite differences in the minimisation techniques employed, results and uncertainties quoted were very similar, and that \textsc{pan} produced more consistent values across the whole wavelength range. We are therefore confident with the quality of results derived by \textsc{pan}.

For a consistency check on the variability of the instrumental profile, we performed Gaussian fits to the unresolved [O\one]$\lambda5577$ telluric emission line on pre-sky-subtracted spectra. The spatial distribution of the fit components clearly show no systematic variability across the IFU field indicating that the line profile shape is very stable. We also find that the telluric line is well fit by a single Gaussian, and shows no evidence for broad wings. The [O\one] line fluxes have a standard deviation, \mbox{s.d.\ $= 1.5\times 10^{19}$~erg s$^{-1}$ cm$^{-2}$ arcsec$^{-2}$}, the FWHM \mbox{s.d.\ $= 4.5$~\kms{}} and radial velocity \mbox{s.d.\ $= 1.9$~\kms}.

We now turn to the uncertainty estimates on each line component's Gaussian properties (line centre, flux and width). Unfortunately, the errors that \textsc{pan} quotes on its fit results  are derived simply from the formal errors on the $\chi^{2}$ minimisation, and we have found these to be an under-estimate of the true uncertainties. More realistic errors have been estimated through visual inspection of the fit quality taking into account noise in the continuum (i.e.~S/N of line) and the number of fitted components. For the H$\alpha$ line flux, the percentage error varies between 0.5--10 per cent for C1 (for high-low S/N lines), 8--15 per cent for C2 and 10--80 per cent for C3. The addition of multiple Gaussian components always increases uncertainties; we estimate that where we can fit a third component, the errors in the flux of C1 and C2 increase by 5--10 per cent. For estimation of the FWHM and radial velocity errors, we have compared fits to the two lines of the [O\three]$\lambda \lambda$4959,5007 doublet where possible. We find the FWHM errors vary between 0.25--2~\kms\ for high-low S/N C1 lines respectively, 4--15~\kms\ for C2 and 15--20~\kms\ for C3. For the radial velocity, the error in C1 varies between 0.1--3~\kms, 2--5~\kms\ for C2 and 10--30~\kms\ for C3. We estimate that the addition of a third component (where required) increases the errors in the C1 and C2 FWHM by $\sim$5~\kms\ and the central velocity by $\sim$3~\kms.

\subsection{Visualisation of the Data Products}\label{daisy}
A major hurdle in the analysis of IFS is the visualisation of the data products. To this end, we have developed a PERL/PGPLOT based plotting and analysis package nicknamed `\textsc{daisy}' that has enabled us to map out the results that we have obtained from our dataset. We have already shown one such visualisation in Fig.~\ref{fig:acs&cont}. The program allows the user to correct for DAR (currently not implemented), correct the measured FWHMs for the intrinsic instrumental profile (via simple FWHM$^2$ subraction), deredden the data using the intrinsic H$\alpha$/H$\beta$ ratio method, convert wavelengths and FWHM into velocities, create a number of ratio combinations, and then to visualise the results. There are options to allow the user to select the colour range (linear or log) for each plot, and to overplot using contours. Once fully developed, our aim is to release this package to the community\footnote{interested users should contact Katrina Exter, kexter@stsci.edu}.

%%%%%%%%%%%%%%%%%%%%%%%%%%%%%%%%
\section{Emission Line Maps} \label{sect:daisyplot}

\begin{figure*}
\includegraphics[width=16cm]{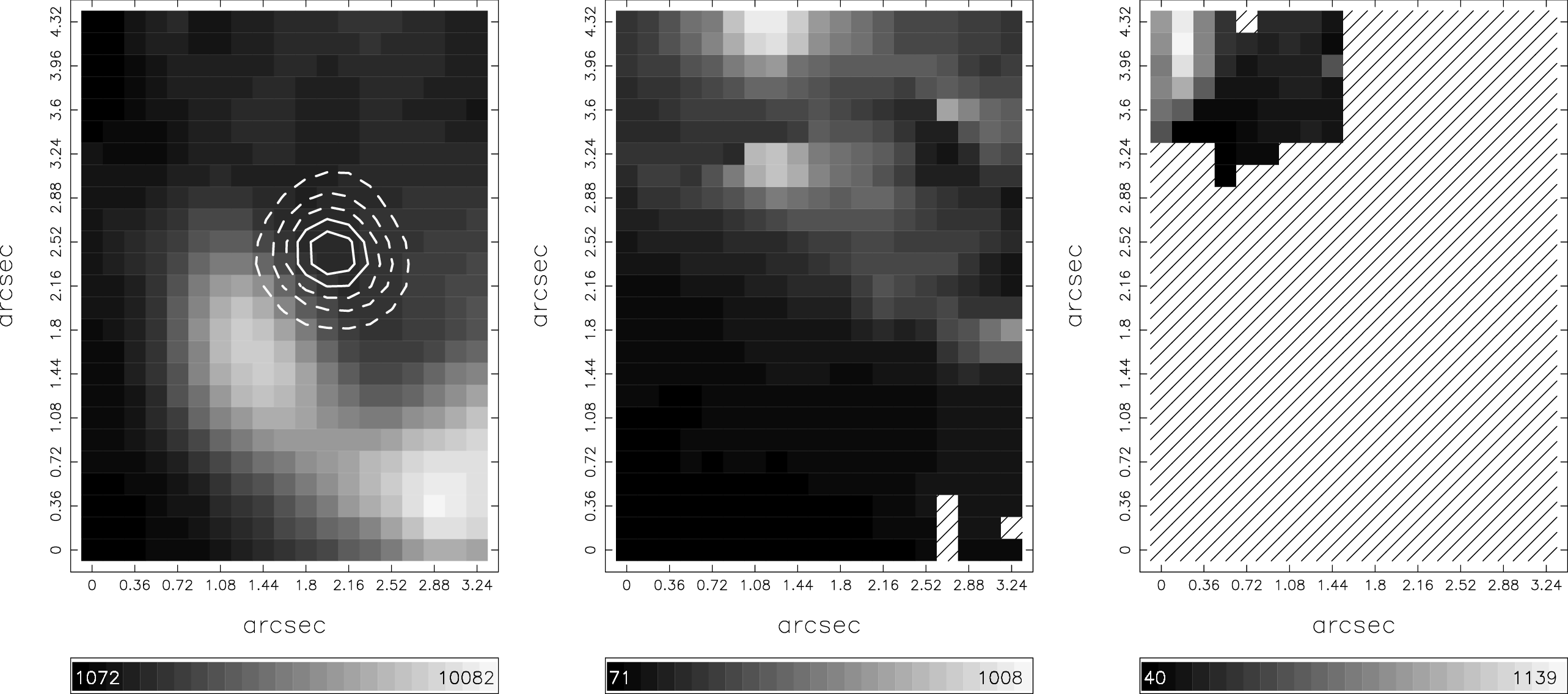}
\caption{\emph{Left:} Flux map in H$\alpha$ component 1 (C1; bright, narrow); \emph{centre:} flux map in H$\alpha$ component 2 (C2; faint, broad); \emph{right:} flux map in H$\alpha$ component 3 (C3; narrow, second peak). Non-detections are represented as hatched spaxels, the $x$ and $y$ scales are in arcseconds offset from the lower-left spaxel, and a scale bar is given for each plot in units of $10^{-15}$ erg s$^{-1}$ cm$^{-2}$~arcsec$^{-1}$. North is up and east is left. The higher levels in the continuum image are plotted as contours on the left-hand plot, identifying the position of cluster 10.}
\label{fig:ha_flux}
\end{figure*}

\begin{figure*}
\includegraphics[width=16cm]{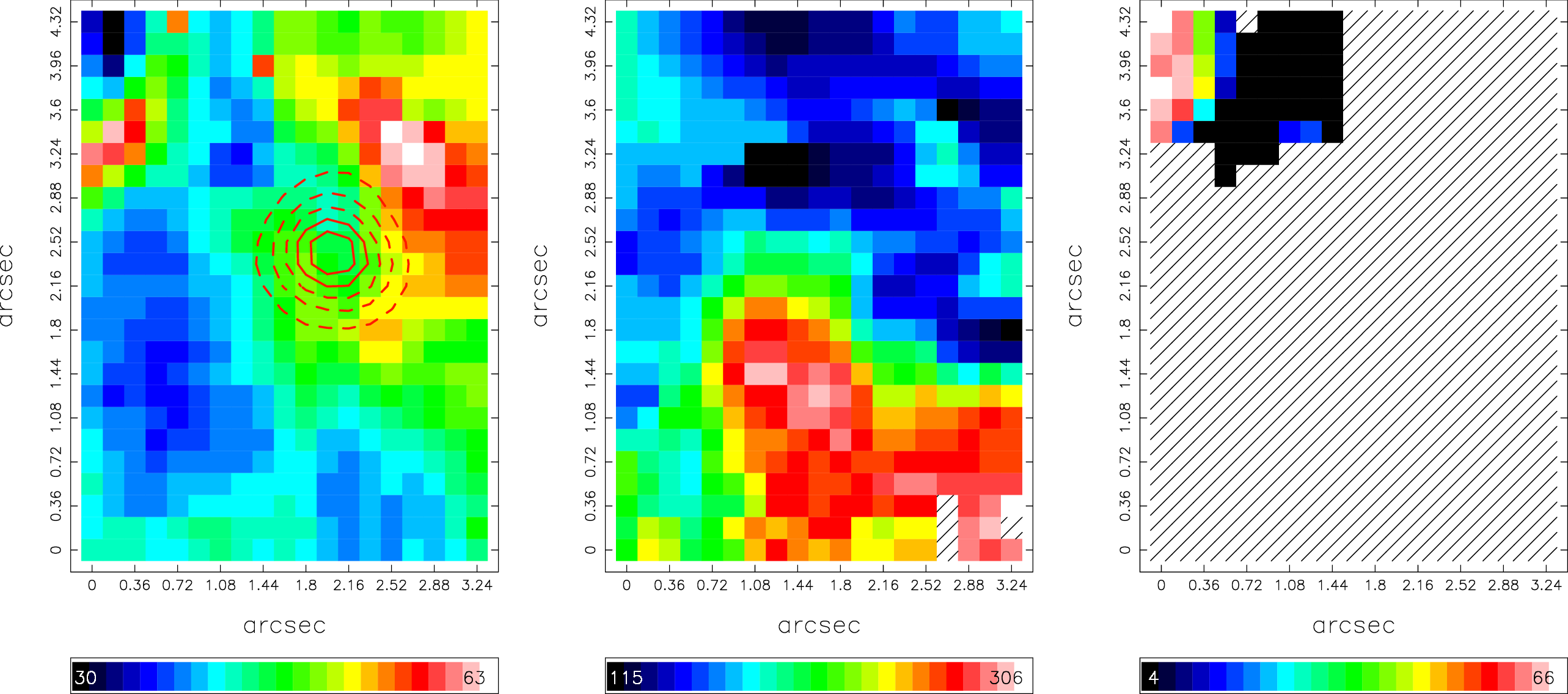}
\caption{\emph{Left:} FWHM map in H$\alpha$ C1; \emph{centre:} FWHM map in H$\alpha$ C2; \emph{right:} FWHM map in H$\alpha$ C3. A scale bar is given for each plot in \kms\ units corrected for instrumental broadening. North is up and east is left, and the position of cluster 10 is identified on the left-hand plot.}
\label{fig:ha_fwhm}
\end{figure*}

Figs.~\ref{fig:ha_flux} to \ref{fig:ha_vel} show maps of the fitted Gaussian properties (flux, FWHM and radial velocity) for each H$\alpha$ line component (we show maps for measurements of the H$\alpha$ line only for the reasons given above in Section~\ref{sect:pan}) over the whole field. The grids represent that of our interpolated, cubed data, not the original GMOS configuration, and the position of the cluster is overlaid with flux contours on selected maps to aid description and interpretation. As described in Section~\ref{sect:pan}, in general we find a bright, narrow component overlaid on fainter, broad emission. This H$\alpha$ broad component reaches a corrected FWHM of $\sim$300~\kms\ in places. In a number of spaxels we have been able to fit a third component (C3), which has approximately the same width as C1 implying that it may be associated with expanding shell- or bubble-like structures in the ionized gas.

\subsection{H$\alpha$ Maps}

\subsubsection{Component 1 (C1)}
From Fig.~\ref{fig:ha_flux}, it is immediately obvious that C1 and C2 do not originate in the same place, i.e.~the flux maps show strikingly different spatial distributions. The two main knots of emission in C1, located near the middle and south-west of the field (south of the cluster), correspond to the pattern of diffuse emission seen in both the WFPC2 F656N (Fig.~\ref{fig:finder}) and ACS F555W (Fig.~\ref{fig:acs&cont}) images. These knots lie just to the south of cluster 10 and mark the easternmost extent of the large ionized complex seen in Fig.~\ref{fig:finder} \citep[H\two\ region 2;][]{waller91}. The width of C1 shown in Fig.~\ref{fig:ha_fwhm} (left) ranges from a FWHM of $\sim$30--40~\kms{} (corrected for instrumental broadening) towards the eastern side of the field, to $\sim$60~\kms{} seen in a region arcing round to the west of cluster 10. This is coincident with the strip in-between two linear filaments seen to the north-west of cluster 10 in Fig.~\ref{fig:finder}, and may be indicative of an influence from the two SNRs located to the north of the FoV. A gradient in FWHM from $\sim$35 to 45~\kms{} is centred on the position of the bright H$\alpha$ knot in the centre of the field, with broader widths seen on the side facing cluster 10, but the line width stays constant to within $\sim$5~\kms{} over the extent of the cluster. As shown by the C1 radial velocity map (Fig.~\ref{fig:ha_vel}, left), we see a region of material moving at $\sim$$-10$~\kms{} in the centre (just to the south of the cluster) surrounded by a ring of material with a velocity decreasing down to the systemic value. There is also an area of gas just to the north moving with a radial speed of $\sim$+7~\kms. It is worth noting that these regions of negative and positive velocities are located either side of the position of the cluster at approximately equal distances. The fastest radially approaching C1 gas is seen in the south-west, moving at 10--15~\kms\ (coincident with the small dense knot seen in the \textit{HST} H$\alpha$ image), and the fastest receding gas is found along the eastern edge, increasing to $\sim$15--20~\kms{} (off the colour scale for this plot) in the far north-east. It should be borne in mind that the total C1 radial velocity range observed across the whole field is only $\sim$35~\kms.

\subsubsection{Component 2 (C2)}
The two small bright knots seen in C2 (Fig.~\ref{fig:ha_flux}, centre) towards the north of the field (north-east of cluster 10) do not correspond to anything seen in the \textit{HST} images. This is understandable since the most intense emission in C1 is an order of magnitude brighter than that in C2. The distribution of C2 FWHMs (Fig.~\ref{fig:ha_fwhm}, centre) is almost precisely anti-correlated with the line flux (Fig.~\ref{fig:ha_flux}, centre), where the brightest features exhibit the narrowest line width (115--125~\kms). The left-hand panel of Fig.~\ref{fig:overplot_maps} better illustrates this by showing contours of C2 flux overlaid on the C2 FWHM map. The broadest C2 lines ($\sim$300~\kms) are seen in the south-centre extending to the south-west of the field, corresponding well with the position of the brightest knots seen in C1 (Fig.~\ref{fig:ha_flux}, left) and the WFPC2 image. This again is illustrated clearly in Fig.~\ref{fig:overplot_maps} (right-hand panel), where contours of C1 flux are overlaid on the C2 FWHM map. The C2 radial velocity field (Fig.~\ref{fig:ha_vel}, centre) covers the same overall range as the C1 gas; material approaching us at velocities of $\sim$15~\kms{} (shown in blue) is located near the northern face of the bright C1 knot in the centre of the field, in-between the negative and positive velocity regions in C1, and almost coincident with cluster 10. This area of approaching gas also extends towards the west, reaching a maximum approaching velocity of 20~\kms. The fastest receding C2 velocities (red) are found along the eastern edge of the field at $\sim$20--25~\kms{} relative to the average. The spatial distribution of this region of redshifted gas mimics that of the redshifted C1 gas, but extends further to the west. Note the distribution of C2 radial velocities does not correlate with either of the C1 or C2 flux maps, however we do tentatively point out a possible radial velocity difference of $\sim$25~\kms{} between the two bright C2 knots seen in the flux map (Fig.~\ref{fig:ha_flux}, centre), and a spatial coincidence with the position of VLA-16 (a possible extended low surface brightness SNR).

\begin{figure*}
\includegraphics[width=16cm]{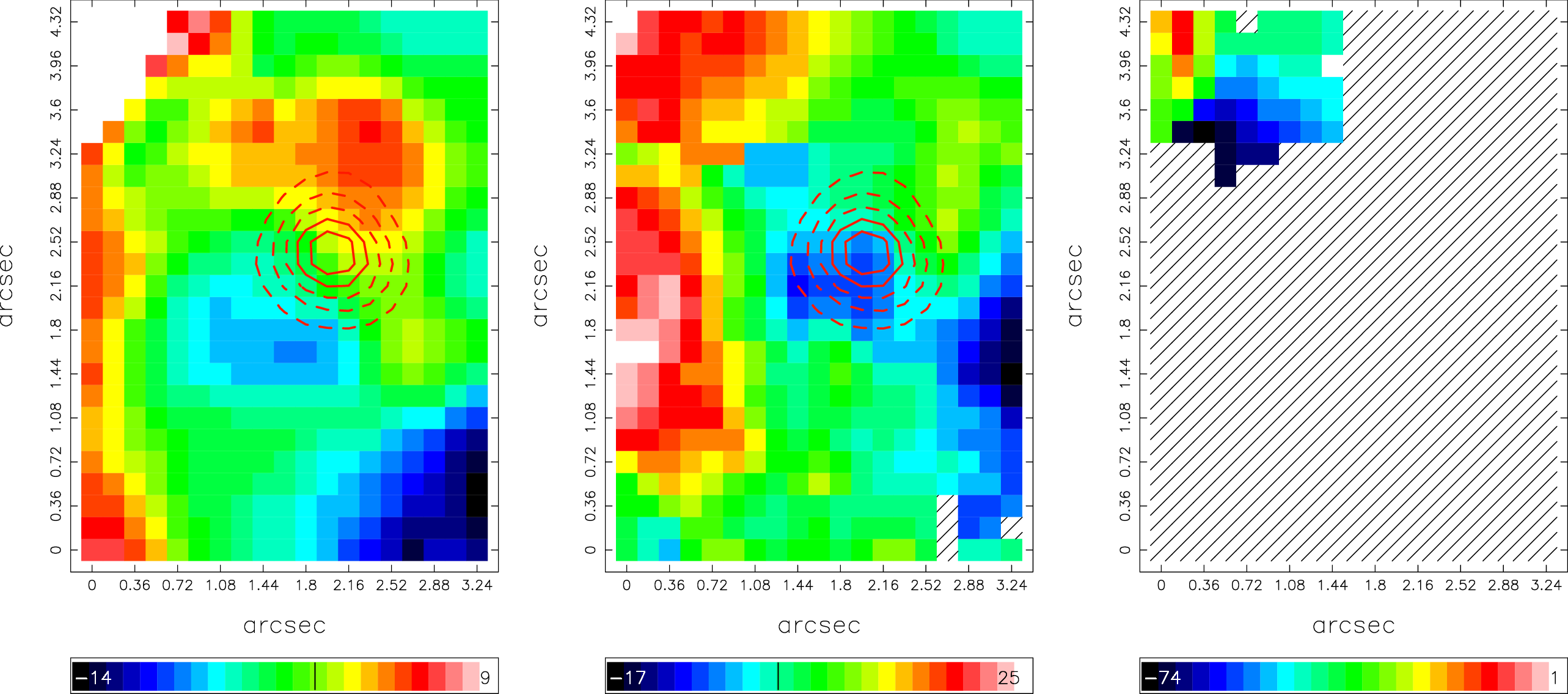}
\caption{\emph{Left:} radial velocity map in H$\alpha$ C1; \emph{centre:} radial velocity map in H$\alpha$ C2; \emph{right:} radial velocity map in H$\alpha$ C3. A scale bar is given for each plot in units of \kms{} (heliocentric) relative to the systemic velocity of the region ($-80$~\kms), where zero is marked with a line. The position of cluster 10 is identified on the left and central plots for comparison.}
\label{fig:ha_vel}
\end{figure*}

\subsubsection{Component 3 (C3)}
A third component to the H$\alpha$ line profile fit was required in a small region located in the north-east of the field. The flux map (Fig.~\ref{fig:ha_flux}, right) shows it is brightest in the far north-eastern corner and has a FWHM range (Fig.~\ref{fig:ha_fwhm}, right) similar to C1. The radial velocity map (Fig.~\ref{fig:ha_vel}, right) indicates that the C3 gas has a larger velocity towards our line-of-sight than any of the C1 or C2 gas (up to $\sim$70~\kms), but the associated uncertainties are large (see Section~\ref{sect:errors}). Since our only detection of this third component is very near the edge of the field, it is difficult to estimate its extent or role from just these maps. Perhaps more interesting is the non-detection of a third component in locations where signatures of expanding structures might be expected from inspection of Fig.~\ref{fig:finder}. We discuss the significance of this component in context with the whole central region and wind-blown cavity in \nocite{westm07b}Paper~II.

\begin{figure*}
\includegraphics[width=11cm]{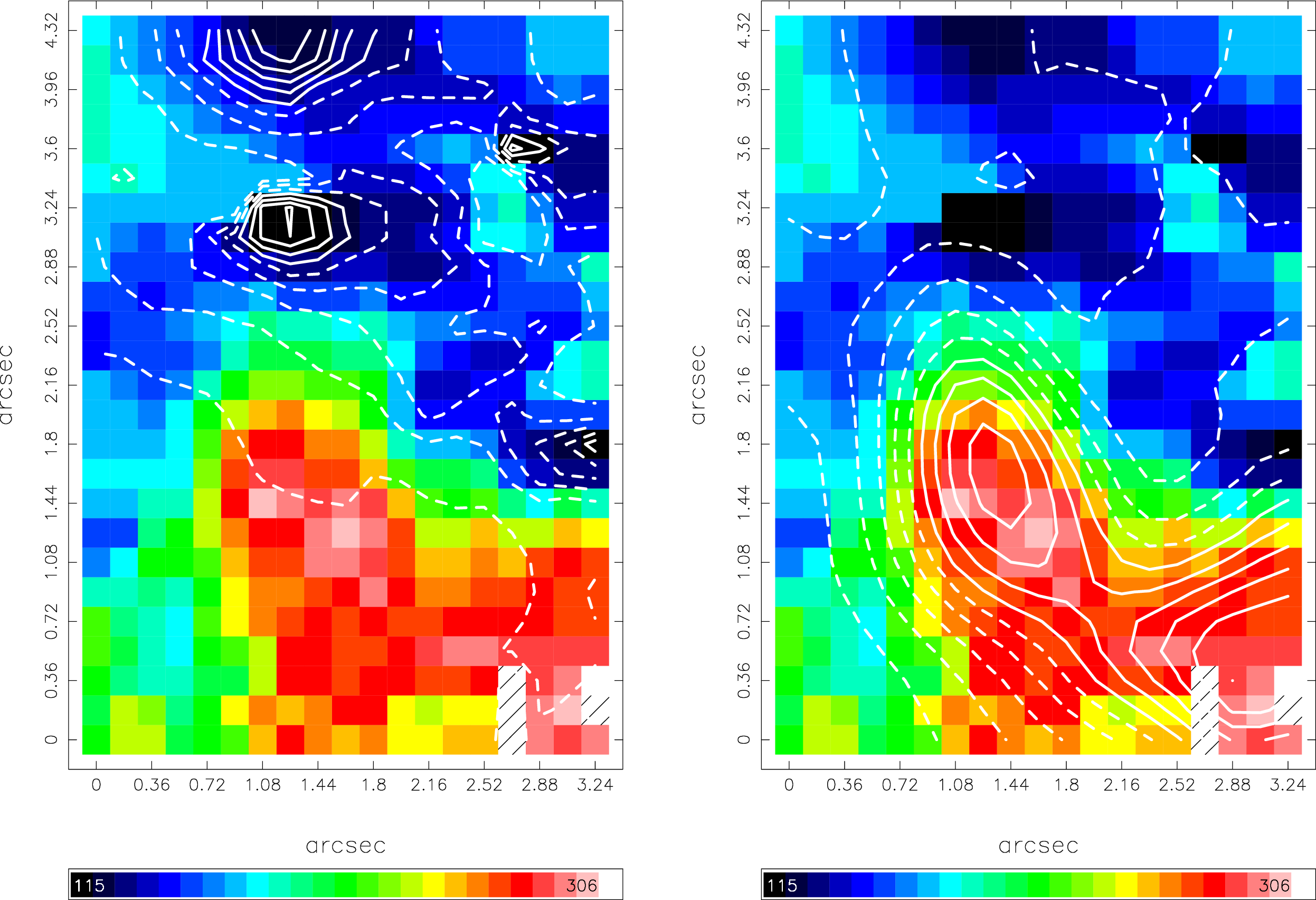}
\caption{\emph{Left:} contours of C2 flux overlaid on the C2 FWHM map. Solid contours represent levels above the mean of the range, and dashed below the mean. Notice the precise correlation between areas of high flux and low FWHM in the north. \emph{Right:} contours of C1 flux overlaid on the C2 FWHM map. Here the correlation between the bright H$\alpha$ knots seen in C1 and the broadest C2 lines can be seen clearly.}
\label{fig:overplot_maps}
\end{figure*}

%%%
\subsection{Flux ratio maps}

\subsubsection{Electron density}
Fig.~\ref{fig:elecdens} shows a plot of the spatial variation of electron density, $n_{\rm e}$, as derived from the ratio between single Gaussian fits to the [S\two]$\lambda\lambda 6713,6731$ lines, assuming an electron temperature, $T_{\rm e}$, of $10^{4}$~K (the S/N of these lines is not high enough to fit multiple line components). There are two distinct regions of higher density seen in the map: in the south-west of the field, $n_{\rm e}$ reaches $\sim$250~cm$^{-3}$ and is spatially coincident with the south-western bright knot seen in H$\alpha$ C1 (Fig. \ref{fig:ha_flux}, left), and the small distinct knot seen in the \textit{HST} F656N image. A second high density region is coincident with the north-eastern part of the central bright H$\alpha$ C1 knot and the position of cluster 10, and peaks at $\sim$270~cm$^{-3}$. We derived the electron density of the H\two\ region surrounding cluster 10 in Section~\ref{sect:clus10}, and found $n_{\rm e} = 175^{+18}_{-20}$~cm$^{-3}$. Our peak value of 270~cm$^{-3}$ is entirely consistent with this value since the spectrum for cluster 10 was created by summing 29 spaxels centred on this central density peak, and hence includes some dilution from the surrounding lower density gas. These values are also consistent with \citet{heckman95}, who find densities of up to $200\pm 100$~cm$^{-3}$ in the central 15$''$ of NGC 1569. Only a small proportion of the gas is at or below the low density limit \citep[$\sim$100~cm$^{-3}$;][shown by the single contour level on the map]{osterbrock89}.

\begin{figure}
\centering
\includegraphics[width=6cm]{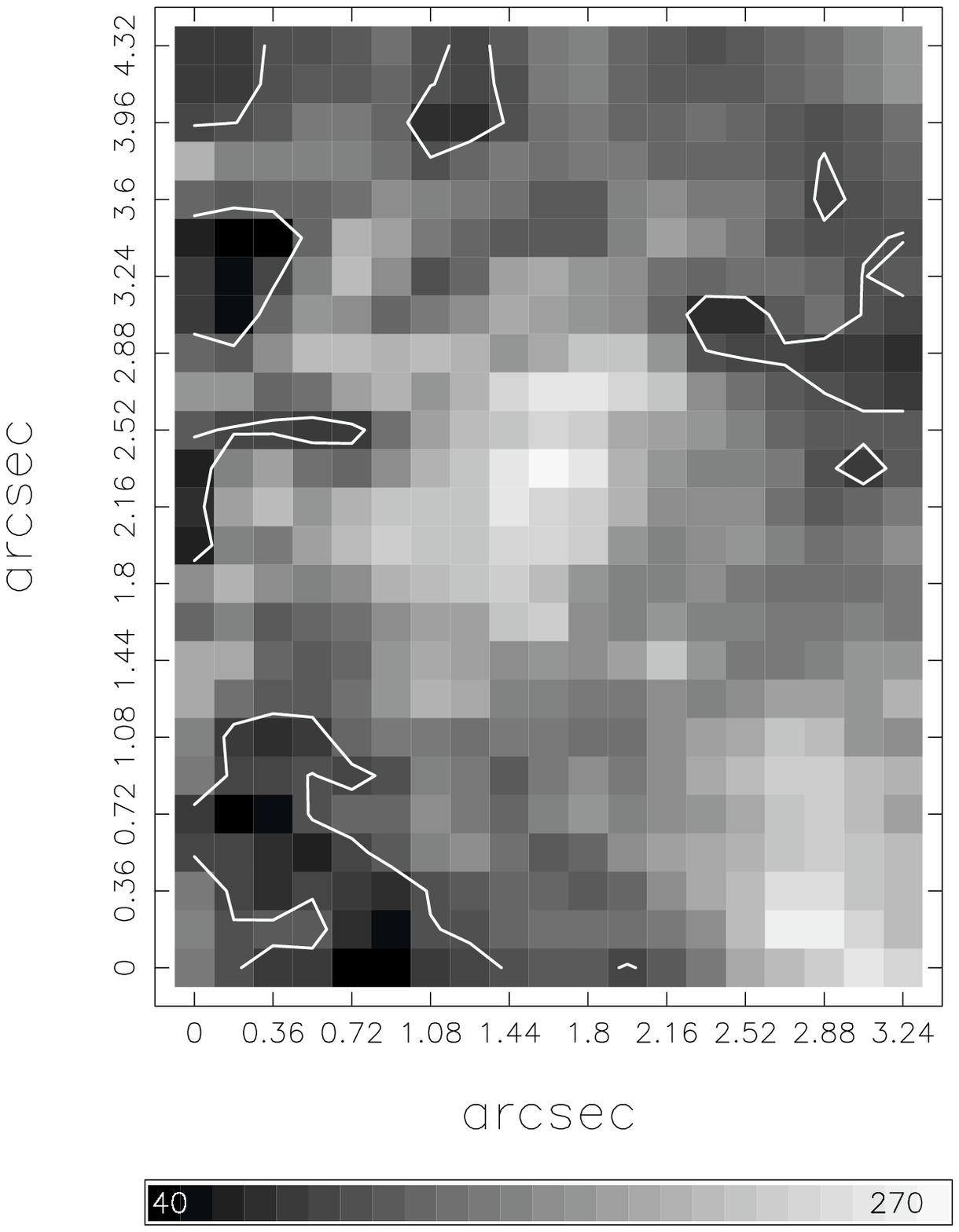}
\caption{Map of the electron density, $n_{\rm e}$, calculated from the ratio of single Gaussian fits to the the [S\two]$\lambda6716,6731$ lines. The scale is in units of cm$^{-3}$. A single contour level marks the 100~cm$^{-3}$ low density limit.} 
\label{fig:elecdens}
\end{figure}

\subsubsection{Nebular diagnostics} \label{sect:diag}
Our wavelength range gives us access to a number of important optical diagnostic lines that can be used to constrain the properties of the emitting gas. The flux ratio of [O\three]/H$\beta$ is a good reddening-free indicator of the mean level of ionization (radiation field strength) and temperature of the gas, whereas [S\two]/H$\alpha$ or [N\two]/H$\alpha$ are indicators of the number of ionizations per unit volume \citep[ionization parameter;][]{veilleux87, dopita00}. A greyscale representation of the observed C1 flux ratio of [O\three]$\lambda 5007$/H$\beta$ overlaid with contours of the [S\two]($\lambda$6717+$\lambda$6731)/H$\alpha$ C1 flux ratio is shown in Fig.~\ref{fig:flux_ratio}. The ratio of $\log$([S\two]/H$\alpha$) is lowest ($-1.7$; shown by dashed contours) where the flux in H$\alpha$ C1 is highest (see Fig.~\ref{fig:ha_flux}, left), i.e.~where the central bright knot is located. The peak of the [O\three]/H$\beta$ ratio is also near to the central bright H$\alpha$ knot, but offset by $\sim$$4\farcs5$ to the south-east. Numerically, the field average ratio of $\log$([S\two]/H$\alpha$) = $-1.29\pm 0.17$, $\log$([N\two]/H$\alpha$) = $-1.62\pm 0.16$, and $\log$([O\three]/H$\beta$) = $0.87\pm 0.05$ (one $\sigma$ errors), showing that the variations in the ratios are really only small. Plotting these values on the standard diagnostic diagrams of the type first proposed by \citet{veilleux87}, shows that all the points lie tightly clustered in the upper-left of the starburst--H\two{} region distribution. Comparing these values with the Mappings-III photoionization models of \citet{dopita00} for an instantaneous burst of star-formation whose SED is given by \textsc{sb99} \citep{leitherer99}, the gas is of low metallicity ($\sim$0.2~\Zsol) and has a high ionization parameter ($q\simeq 10^{8}$), as expected. \citet{buckalew06} use \textit{HST}/WFPC2 narrow-band imaging to calculate the ratios of [O\three]/H$\beta$ and [S\two]/H$\alpha$, and use the theoretical `maximum starburst line' derived by \citet{kewley01} to set a threshold to determine the difference between photoionized and non-photoionized emission. Using this criterion on  our de-reddened line ratios, we find a total of 118 spaxels containing possible non-photoionized emission, the location of which are plotted on Fig.~\ref{fig:flux_ratio} with white crosses. We note here that the threshold derived by \citeauthor{kewley01} is a maximum in the sense that no pure starburst, regardless of metallicity (in the range 0.1--3~\Zsol) or ionization parameter, could produce emission above this line. A lower metallicity, such as found in NGC 1569, decreases the level of this theoretical limit, but since the model predictions are quite uncertain, understanding how many more points may have a non-photoionized contribution to their emission is non-trivial. For this reason, we have employed the more robust `maximum' threshold.

The distribution shown in the map corresponds well to the maps of \citet{buckalew06}, with the majority of points lying in the south-east of the field. Since SSC A is located only $7''$ away in this direction, \citet{buckalew06} associate these points with an `arc' of non-photoionized gas resulting from a possible wind--wind interaction between the two clusters. %However, since our derived ratios only just fall above the maximum starburst line, we cannot be sure that non-photoionization processes play a significant part in the observed ionization levels \citep[see][]{calzetti04}.

\begin{figure}
\centering
\includegraphics[width=6cm]{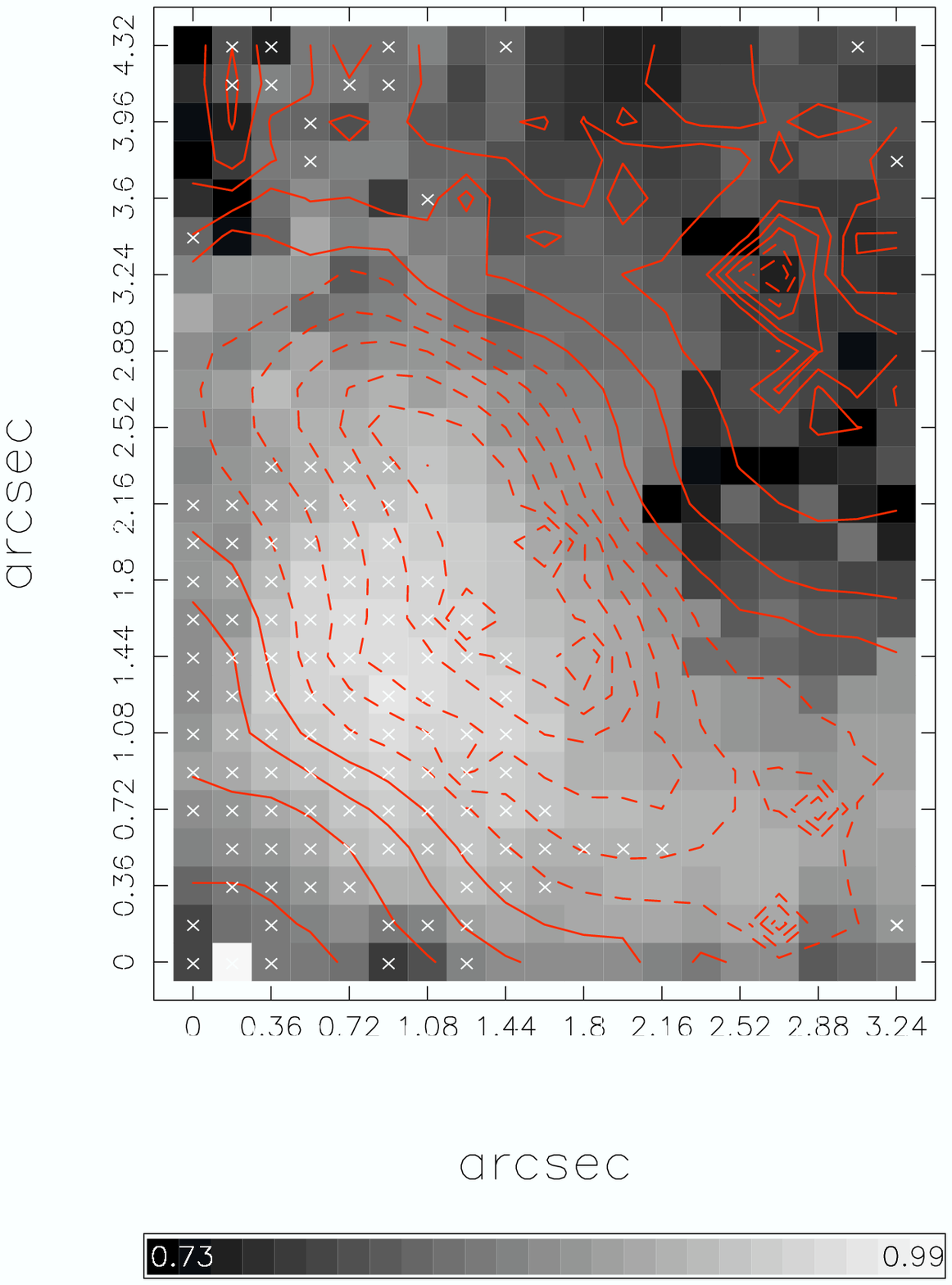}
\caption{Flux ratio of $\log$([O\three]$\lambda 5007$/H$\beta$) C1 (greyscale; scale bar shown) with contours of the ratio $\log$([S\two]($\lambda$6717+$\lambda$6731)/H$\alpha$) C1 (12 linearly spaced levels from $-1.7$ to $-0.9$). Spaxels marked with a cross indicate non-photoionized emission according to the maximum starburst line of \citet{kewley01}.}
\label{fig:flux_ratio}
\end{figure}

\subsubsection{Reddening map}
De-reddening with our visualisation and analysis package uses the Galactic law of \citet[][$R_{V}=3.1$]{howarth83} and the intrinsic H\one{} ratio of H$\alpha$/H$\beta$ = 2.85 \citep[][for gas at $T_{\rm e}$ = $10^{4}$~K and $n_{\rm e}$ = 100~cm$^{-3}$]{storey95}. A reddening map, computed using only C1 of H$\alpha$ and H$\beta$, is shown in Fig.~\ref{fig:chb}. The mean and standard deviation for c(H$\beta$) for C1 over the field-of-view is $1.01\pm0.16$ (compared to 1.15 when summing the flux in all components for the H\one{} lines). In this, we are assuming most of the reddening comes from a Galactic foreground screen \citep[a valid assumption;][Section~\ref{sect:clus10}]{israel88, origlia01}. The reddening distribution is consistent with the location of the central and south-western bright knots seen in C1, where these knots exhibit the highest extinction [c(H$\beta$) = 1.3 $\equiv$ $E(B-V) = 0.8$]. Interestingly, high values are also found in the north-west of the field, not correlating with any feature on any of the other maps. The lowest reddening [c(H$\beta$) = 0.75 $\equiv$ $E(B-V) = 0.5$; equivalent to the Galactic foreground level; \citealt{relano06}] is found in the north-east of the field, coincident with the location of the group of spaxels containing C3.
%The absolute fluxes, not just the slope of the spectra, are altered by the dereddening.   

\begin{figure}
\centering
\includegraphics[width=5.5cm]{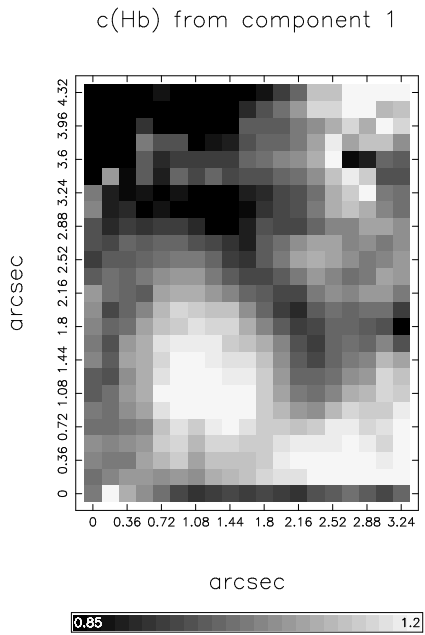}
\caption{A c(H$\beta$) map calculated from the C1 flux ratio of H$\beta$/H$\alpha$.}
\label{fig:chb}
\end{figure}

%%%%%%%%%%%%%%%%%%%%%%%%%%%%%%%%%
\section{Interpretations and Discussion}\label{discussion}

At the distance of NGC 1569, the GMOS IFU field-of-view (FoV) covers $\sim$$50\times 35$~pc with an 8~pc resolution, and is therefore ideal for studying the detailed relationship between young star clusters and their environment. To this end, we have observed a region near the centre of the galaxy covering one such young, bright star cluster and the surrounding ionized gas. Taking advantage of the spatially resolved nature of IFU observations, we extracted a spectrum of the cluster to derive its fundamental properties, and created well-defined, high S/N maps of the Gaussian properties of each observed H$\alpha$ line component across the FoV. In the following sections we discuss our findings relating to the cluster itself and to the surrounding ionized gas. We also explore possible theories that might give rise to the observed broad emission profiles and discuss their consequences.

\subsection{Cluster 10}
As clearly seen in the high-resolution \textit{HST}/ACS images (Fig.~\ref{fig:acs&cont}), cluster 10 is composed of two individual clusters which we denote 10A and 10B. We measure their projected separation to be $\sim$3.7~pc and their sizes to be $R_{\rm eff, A} = 0.88$ and $R_{\rm eff, B} = 0.60$~pc, and a combined photometric mass of 7$\pm$$5\times 10^{3}$~\Msol. By comparing the relative photometry of each cluster we find them to have very similar ages: 10B is the youngest ($<$5~Myr) and 10A a little older (5--7~Myr). The combined spectrum of the two clusters extracted from our IFU data shows that one (or both) of these clusters has a significant population of WR stars.

The unusually compact nature of these two clusters is reminiscent of R136 in the LMC. This equally young cluster has an $R_{\rm eff} \approx 1.7$~pc and mass of a few\,$\times$\,$10^{4}$~\Msol{} \citep{hunter95, bosch01}. R136 contains a large number of massive stars that power and maintain the $\sim$100~pc-diameter giant H\two{} region 30 Dor.

Our age estimates of the two clusters show that their formation was approximately coeval with SSC A, and that vigourous star-formation has certainly occurred recently in NGC 1569. Cluster 10 is located within the bright H\two{} region No.\ 2 \citep{waller91} which is found to emit a powerful thermal radio signature \citep{greve02} and bright mid-IR forbidden-line and continuum emission \citep{tokura06}. These observations imply that this whole area is a large embedded star-formation complex being fuelled by the neighbouring molecular CO cloud \citep{taylor99}.

We do not see evidence for a compact individual H\two{} region surrounding the cluster in the H$\alpha$ flux maps or the extinction map. This is probably because the winds and supernova events have blown away the surrounding gas. The electron density map, however, does show an enhancement at the location of cluster 10, indicating that perhaps there is some associated gas remaining. Our IFU FoV contains a great deal of ionized material that is both influenced by cluster 10 and by the relatively distant, but more massive, SSCs A and B. It is not clear if the source of ionization for H\two{} region 2 is solely cluster 10 or whether other starburst regions contribute.

\subsection{Ionized gas} \label{sect:1569paperI_gas}
\subsubsection{The turbulent ISM and the narrow component (C1)}

The integrated H\one{} velocity dispersion for the disc of NGC 1569 is $\sigma \approx 15$~\kms{} (FWHM $\approx$ 35~\kms) indicating that the neutral ISM is very disturbed \citep{muhle05}, presumably due to the combined effects of stellar winds and supernovae. This value is consistent with the broad line widths also found in H\one{} clouds associated with nearby giant H\two{} regions \citep*{viallefond81}. We would thus expect the H\two{} gas in NGC 1569 to have similar or higher velocity dispersions to the H\one{}.

In Fig.~\ref{fig:sig_vel} we plot the heliocentric radial velocity against line width for all identified line components across the FoV, whilst in Fig.~\ref{fig:sig_flux} we plot the line flux against line width. This latter plot was used successfully as a diagnostic tool by \citet[][see also \citealt{martinez07}]{m-t96} to examine the physical origin of the broadening mechanisms in giant H\two{} regions. Plotting the line characteristics in this way highlights a number of trends within and between the components.

\begin{figure}
\includegraphics[width=0.47\textwidth]{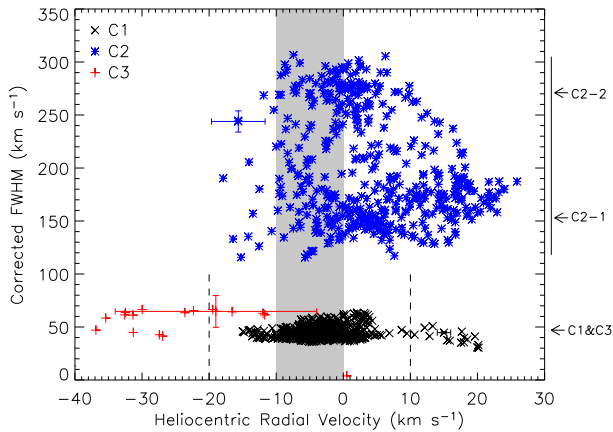}
\caption{H$\alpha$ line width (FWHM, corrected for instrumental broadening but not thermal broadening of $\sim$20~\kms) vs.\ centroid velocity (heliocentric, relative to the systemic velocity of the region, $v_{\rm sys}=-80$~\kms). A representative error bar has been plotted for each of the three components on one selected data point. The shaded region represents the velocity (including the error) of cluster 10 (see Section~\ref{sect:clus10}). Next to the right-hand axis are labels marking the three regimes in FWHM discussed in the text. C1 (C3) points lying to the right (left) of the vertical dashed lines form the high-velocity tails referred to in the text.}
\label{fig:sig_vel}
\end{figure}

\begin{figure}
\includegraphics[width=0.47\textwidth]{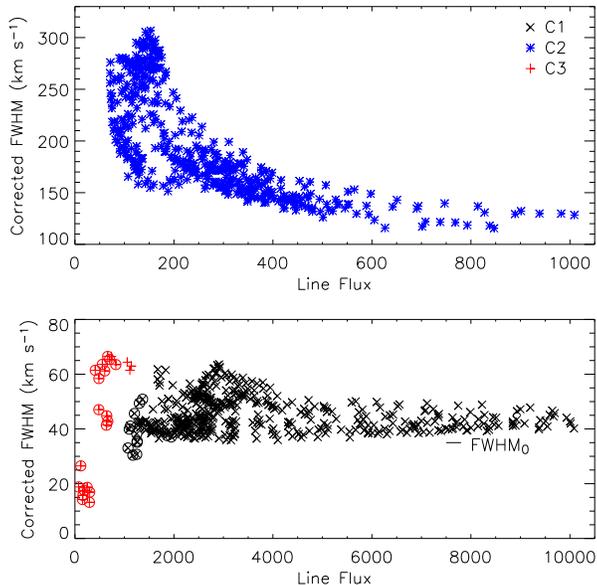}
\caption{H$\alpha$ line width (FWHM, corrected for instrumental but not thermal broadening) vs.\ flux (arbitrary scale). The selected axis ranges show C1 and C3 in the lower panel, and C2 in the upper panel (note in particular the factor 10 difference in flux range). The distinct lower limit to the FWHM of C1, FWHM$_{0} \approx 35$~\kms{}, is marked. Encircled points are those belonging to the high-velocity tails referred to in the text and defined by the vertical dashed lines in Fig.~\ref{fig:sig_vel}.}
\label{fig:sig_flux}
\end{figure}

Fig.~\ref{fig:sig_vel} shows that components C1 and C3 (FWHM = 35--65~\kms) are easily distinguished from component C2 (FWHM = 120--300~\kms). Furthermore, in the bottom panel of Fig.~\ref{fig:sig_flux}, a distinct $\sim$35~\kms{} lower limit to the C1 FWHM (which we henceforth refer to as FWHM$_{0}$) can be seen. After correction for thermal broadening ($\sim$20~\kms at $T\sim 10^{4}$~K), this becomes 29~\kms{} (or $\sigma \approx 12$~\kms), meaning the C1 line velocity widths correspond to mildly supersonic speeds (for $T_{\rm gas}=10^{4}$~K, the sound-speed is $\sim$10~\kms). We note that the C1 line width does not correlate strongly with line flux, unlike the broad component (C2) which we discuss below.

The red velocity tail of C1, seen clearly on Fig.~\ref{fig:sig_vel} as points lying to the right-hand side of the vertical dashed line at velocity = +10~\kms, corresponds spatially to the extreme north-east of the field where we also see a blueshifted component (C3; see e.g.~Fig.~\ref{fig:fitegs}, lower panels). These C3 points lie to the left-hand side of dashed line at velocity = $-20$~\kms. The velocities of these double-peaked emission lines are distributed evenly to the red and blue of the cluster radial velocity (shown as a grey strip), providing the only signature of classical shell expansion in the field. Furthermore, the two components have similar flux levels as illustrated in Fig.~\ref{fig:sig_flux} (see also Fig.~\ref{fig:fitegs}, lower-right panel), where the corresponding high-velocity C1 and C3 points are plotted with encircled symbols.

Broad line widths have been observed in many young star-forming regions e.g.~30 Dor or NGC 604 \citep{chuken94, yang96}. A search through the literature uncovers a number of alternative theories that could account for the observed line-widths of C1, including electron scattering, multiple unresolved expanding shells integrated along the line-of-sight, and gravitational broadening through virial motions.

Electron scattering can be immediately ruled out, as unphysical electron densities are required to produce the observed widths \citep[see][]{roy92}. Multiple unresolved expanding shells integrated along the line-of-sight was an explanation proposed by \citet{chuken94} to explain their observed broad line-widths in the giant H\two{} region 30 Dor. A follow up study of the same region by \citet*{melnick99}, but at very high spatial- and spectral-resolution ($0\farcs6$ seeing and 0.067~\Apix), allowed shells to be resolved down to their resolution limit of 0.13~pc. In their observations, the broad line-widths found by \citealt{chuken94} were resolved into a myriad discrete narrow lines representing individual shell components, thus validating the conclusions of \citet{chuken94} when observations are made at insufficient spatial/spectral-resolution.

However an analytical model for this unresolved shells effect, first developed by \citet{dyson79} and analysed in detail by \citet*{t-t96}, predicts that if the age distribution of the stars powering the expanding shells is greater than the timescale for gravitational collapse of the parent cloud, the resulting line profiles are always flat-topped. This is something we do not see. Since the spatial and spectral resolution of our observations is $\sim$9~pc and 0.34~\Apix, it is conceivable that a contribution to the narrow-line width is due to unresolved expanding shells. However, since the point at which coherent shells break up is when their expansion velocity equals the ambient ISM turbulent velocity (FWHM$_{0}$), unresolved expanding shells cannot contribute significantly to the observed widths of C1.

Furthermore, \citet{t-t96} find that to have a collection of unresolved shells capable of producing a supersonic line-width over a large range of densities, implies a great many low-energy winds with properties similar to those typical for low-mass stars. If multiple shells are responsible for the observed broadening in our case, then the level of broadening would be expected to scale to a certain degree with gas column density (a large volume of gas could contain shells with a larger range of velocities), but this is not observed. The broadest C1 lines are seen in the north-west of the field, away from the bright knots in the field centre.

Gravitational broadening has been suggested by a number of authors as an explanation for the observed line-widths in giant H\two{} regions (\citealt{melnick77, terlevich81}; \citealt*{t-t93}), and predicts that the line profile should be approximated by $\sigma_{\rm virial} = \sqrt{GM/R}$. This explanation has been successfully used to explain and calibrate the $R\propto \sigma$ and $L\propto \sigma$ relations found for giant H\two{} regions \citep*{terlevich81, melnick88}, and explain the supersonic widths found in NGC 604 \citep{m-t96, yang96}. Although NGC 1569 cannot be likened to a single giant H\two{} region, we can make a crude estimate of the expected virial broadening by taking the mass to be the total H\one{} mass of the galaxy \citep[$M_{\rm HI} \sim 10^{8}$~\Msol;][]{muhle05} and an isophotal radius of $\sim$4~kpc \citep{vaucouleurs91, muhle05}. This results in a broadening of $\sigma_{\rm virial} \approx 11$~\kms, in surprising agreement with our measured FWHM$_{0}$.

%However it is unlikely that the stellar motions could be transmitted to the gas with enough efficiency that this effect could dominate over the general turbulence resulting from the intense star-formation episodes experienced by the galaxy.

Thus, in conclusion we find that C1 represents the turbulent ISM, where its motions result from a convolution of both the underlying gravitationally induced virial motions inherent in the gas and the general stirring effects of the starburst. It is also possible that an additional minor contribution to its width may originate from unresolved shell components along the line-of-sight.

\subsubsection{The broad component, C2}
We now turn our attention to the intriguing broad component (C2). In the following discussion, we will attempt to understand what the conditions of the emitting gas are, and what mechanism(s) are acting to give it such a broad width. Firstly, we can use Figs~\ref{fig:sig_vel} and \ref{fig:sig_flux} to identify a number of trends to help constrain the physical properties of this component. Fig.~\ref{fig:sig_vel} shows that the group of C2 points can be split into two further sub-groups. The narrower group (FWHM\,$\approx$\,150~\kms; labelled C2-1 on the plot) is associated with the regions of higher C2 flux (compare the central panels of Fig.~\ref{fig:ha_flux} with Fig.~\ref{fig:ha_fwhm}), whereas the close grouping of high-FWHM C2 measurements ($\approx$\,280~\kms; labelled C2-2) correspond (to a high degree of correlation) with the most intense C1 emission (i.e.~the two bright knots; see Fig.~\ref{fig:overplot_maps}, right panel). Disregarding the red tail of the C1 velocity distribution discussed earlier, there is an average redward offset of only $\sim$10~\kms{} between C1 and C2, implying that C2 does not originate from gas with large-scale bulk motions. Fig.~\ref{fig:sig_flux} clearly demonstrates that the broadest C2 lines are also the faintest \citep[a common result;][]{m-t96}, where the average ratio of flux(C2)/flux(C1) over the FoV is 0.07. We can now assert that any theory explaining the origin of C2 must describe both how it exists over the entire FoV, and why its width is at once so highly correlated with the position of the bright knots seen in the narrow component, and so anti-correlated with its own flux.

Mechanisms proposed in the literature that can account for line widths of $>$100~\kms{} in H\two\ regions are quite varied, particularly from studies of more distant systems. Moreover, there are few studies that compare in terms of both the spatial-scale of the overall region (i.e.~30 Dor, for example, is clearly much smaller than the NGC 1569 starburst) and the spatial-resolution of the observations. This makes it harder to evaluate which mechanisms may be at work in NGC 1569.

Unresolved expanding shells have again been cited as a possible cause of the broad component in a few studies, but it is difficult to envisage how smooth Gaussian profiles with widths of up to 300~\kms{} could be fully accounted for in this manner without showing some signs of velocity splitting at our level of spectral resolution. \citet{heckman95} observed H$\alpha$ emission from NGC 1569 using two perpendicularly oriented long-sits intersecting at SSC A, and in positions where they were able to fit multiple Gaussians to the line profile, they found two distinct kinematic systems. Lines near to $v_{\rm sys}$ were found to have widths of FWHM $\approx$ 30--90~\kms, whereas broader lines with widths of up to 150~\kms{} were only identified at radial velocities of over 200~\kms{} relative to $v_{\rm sys}$. \citeauthor{heckman95} proposed that this second, high-velocity system is the origin of the broad widths (30--50~\AA{} at full-width zero-intensity) found in the summed H$\alpha$ line profile of the area surrounding SSC A, since when integrating a larger spatial area, the discrete kinematic systems became blended together\footnote{In our data, we associate C3 with part of the high-velocity system described by \citeauthor{heckman95}, and where detected, we resolve it at a sufficient level for it not to be confused with the broad C2.}. The level of broadening due to unresolved components is therefore dependent on \emph{both} spatial- and spectral-resolution. This is illustrated well by the study of \citet{melnick99}, described above. Although their high spatial- and spectral-resolution observations allowed them to resolve individual line components (shells) on scales down to 0.1~pc, they were still able to identify a clear, underlying, broad component persisting in all sight lines. 

Can the effects of high-energy photons and fast-flowing winds from the nearby clusters explain the origin of C2? Over the years, much work has been done examining these effects using complex hydrodynamical models (\citealt*{hartquist92, klein94}; \citealt{suchkov96, pittard05, marcolini05, t-t06}). An exchange of energy and mass between the different hot and cold phases of the ISM can occur through four main processes: photoevaporation; conductively-driven thermal evaporation; hydrodynamic ablation; and turbulent mixing \citep[for a review see][]{pittard06}. Photoevaporation takes place in the presence of a strong ionizing radiation field, and conductive evaporation occurs when energetic electrons from a hot, surrounding medium, deposit their energy in the cloud surface. When the hot, tenuous medium is fast-flowing, as in the case of stellar or cluster winds, shearing between the high velocity flow and the cloud surface can set up a turbulent mixing layer that can drive further evaporation. The impacting wind can also physically strip (ablate) material off the clump surface, and entrain it into the flow \citep{scalo87}. 

\citet{bf90} described a model for a turbulent mixing layer that forms between the hot, inter-cloud gas and the surface layers of cool or warm clouds in near-pressure equilibrium. The UV radiation emitted from the mixing layer photoionizes the cold gas in the cloud surface giving rise to optical emission lines (in a similar way to a shock precursor) with characteristic velocities of a fraction of the hot-phase sound speed. \citet*{slavin93} expanded on this idea by modelling the cloud interface in detail. Including the effects of nonequilibrium ionization and self-photoionization allowed them to predict a number of optical, IR and UV line ratios, which they found to be in good agreement with the observed values at high latitudes in our Galaxy (for which this model was originally developed). \citet{slavin93} show a schematic of their model in their figure 1: hot gas flows past a sheet of cold or warm gas inducing Kelvin-Helmholtz instabilities and the build-up of a turbulent mixing layer at the interface region. Thermal conduction between the layers results in a turbulent cascade down to the dissipative level, allowing the efficient exchange of energy between the media. Their model predicts strong optical [O\three] and H$\alpha$ emission with non-Gaussian (broad-winged) line profiles caused by the high levels of turbulence. Strong far-UV emission from C\four{} and O\six{} is also predicted. These predictions are qualitatively in good agreement with what we observe, however, the \citeauthor{slavin93} model predicts [O\three]/H$\beta$ and [S\two]/H$\alpha$ nebular line ratios firmly in the shocked regime, far away from our observed ratios. A decrease in the 'maximum starburst' line threshold (indicating gas with a shocked emission component) due to the low metallicity of NGC 1569 may account for part of this discrepancy.

It is interesting to note that observations of CO \citep[e.g.][]{falgarone90}, CH$^{+}$ \citep*{crane91} and H$_{2}$\,2.1\,$\upmu$m \citep[e.g.][]{geballe86} in close-by Galactic molecular clouds have also revealed broad line profiles similar in shape to what we observe in our optical recombination and forbidden lines. To explain these observations, \citet{hartquist92} and \citet{dyson95} also developed turbulent boundary layer models, and found these broad molecular line widths could again be reproduced through the presence of shear-induced turbulence, resulting from shocks in the cloud interface due to the viscous coupling between the fast wind and the clump gas. \citet{melnick99} were the first to suggest a connection with turbulent layers and erosion of dense gas clumps in an extragalactic environment when discussing the persistent underlying broad line component identified from their high-resultion observations of 30 Dor.

We therefore find the most likely explanation of the origin of C2 is emission from a turbulent mixing layer formed at the surface of dense clouds resulting from the viscous coupling between the cool clump material and the hot, fast, winds from the surrounding clusters. Material is also removed from this layer through thermal evaporation and/or mechanical ablation, resulting in the mass-loading of the flow. We know from X-ray studies that the hot, diffuse medium in NGC 1569 has a characteristic temperature around 0.7~keV \citep[$\sim$$10^{7}$~K;][]{martin02}, implying that the sound speed in this hot phase is $\sim$500~\kms. The line-widths we measure are only 60 per cent this value, meaning that the turbulent motions are subsonic with respect to the hot gas. This may provide a further explanation for the lack of shock-excited line ratios in our data, since it is expected that photoionization in these central regions would overwhelm the signature of shock excitation anyway.

Although a shear-induced turbulent mixing layer can theoretically attain a quasi-steady state when the energy flux into the layer is equal to the cooling rate per unit area \citep{slavin93}, the additional destructive effects of evaporation and/or ablation mean that gas clumps will only have a limited lifetime. How long could we therefore expect the clouds to survive? Unfortunately our knowledge of cloud disintegration is not yet developed enough to answer this question in detail. Simple hydrodynamical simulations of ablation predict that clouds enveloped in a supersonic (relative to the cloud temperature) wind have survival timescales typically less than 1~Myr \citep{klein94, marcolini05}. Clearly this cannot be the case since warm ionized gas is observed in superwinds out to heights of many kpc \citep[e.g.][]{shopbell98, martin98, db99}. Further detailed modelling work is required.

Whether or not these mechanisms are all acting in our particular FoV, we conclude that the energetics are dominated by the interface layer at the surface of the gas clumps, and that the principle influence on the gas state is external to the clouds, rather than embedded within the cloud, as would be the case for isolated giant H\two{} regions. In other words, the properties of the H$\alpha$ knots in our FoV are tied to the intrinsic state of the gas rather than the available energy supplies (the star clusters). This important difference may help to explain why the turbulent widths decrease away from the cloud surfaces. Cloud material remains intact and cool only for a limited length of time after it leaves the surface of the gas clump. Dense gas found near the surface of the cloud would be able to emit in H$\alpha$ at higher velocities as it becomes entrained into the high-velocity flow than less dense gas, resulting in a broader line-width at the location of the cloud.

\subsubsection{The motion of C2}
The average width of C2 is much higher than the average radial velocity offset between C1 and C2 (obvious from comparing Figs~\ref{fig:sig_vel} and \ref{fig:sig_flux}), meaning that in our FoV, turbulent motions dominate over bulk flows. \citet{t-t06} develop 2D hydrodynamical models of the evolution of an H\two{} region assuming that the medium surrounding the ionizing star cluster is composed of a large collection of clouds (i.e.~a clumpy medium). Although their model is not specifically applicable to our case since the ionizing source is located within the cloud stratum, it illustrates a number of interesting results regarding the flow of material within the outer bounding shock front. In their simulations, the leading shock filters slowly through the gas clouds dissipating them one-by-one, leaving a highly turbulent, mass-loaded medium in its wake. The material interior to the leading shock moves in all directions resulting from the obstruction by the gas clumps, and possesses only a small net velocity in the outward direction. This is a very similar situation to what we observe, and shows that our FoV must be sampling gas well within the bounding shock (or sonic point) of the galactic wind, since as the flow becomes supersonic, turbulence is damped leaving only the bulk velocity component.

\subsubsection{Final remarks}

We find small-scale variations in the velocity fields of both C1 and C2 components, suggesting local energy inputs. However, these cannot (yet) be tied unambiguously to the effects of any single energy source (e.g.~cluster 10). The distribution of non-photoionized emission covering the whole of the south-eastern portion of the FoV (see Fig.~\ref{fig:flux_ratio}) would suggest that high-velocity gas is impacting from the south-east (i.e.~the direction of SSC A), but without information on the three-dimentional distribution of the gas and velocity field, we cannot be sure.

Since the constraints to our model of the origin of C2 are only qualitative, further observations and modelling are clearly required. We predict that if the energetics are dominated by the interface region where the turbulence is proposed to originate, observations of emission originating from deep within the cloud would not show this line broadening. This would also apply to regions shielded from the wind impact, where only the ambient ISM turbulence would be evident. Furthermore, the models of \citet{marcolini05} predict strong soft X-ray emission from the bow-shock (reverse shock) produced as the wind hits the cloud. Whether the X-rays are primarily emitted from wind material or evaporated cloud material depends on the metallicity of the wind and the strength of the thermal conduction within the cloud. UV emission (or absorption) from C\four{} and O\six{} is also predicted to arise through collisional processes within the cloud material at the wind--cloud interface by both \citeauthor{slavin93} and \citeauthor{marcolini05} Currently, O\six-emitting gas is the highest temperature phase of an outflow that can be probed for which reliable velocities can also be measured \citep{heckman01}, making this transition a very important tool in the investigation of galactic winds.

In a companion paper \nocite{westm07b}(Westmoquette et al. 2007a; Paper II), we present the analysis of the other three IFU fields covering the central regions of NGC 1569. Following on from our analysis presented here, we first concentrate on the details of each field, then combine all the data in an attempt to constrain the general state of the ISM in the whole region.

%%%%%%%%%%%%%%%%%%%%%%%%%%%%%%%%%%
\section{Summary}\label{conc}
The unique capabilities of the GMOS IFU, coupled with \textit{HST} imaging, have allowed us to map the detailed ionized gas properties of a $5 \times 3.5$~arcsecs region near the nucleus of NGC 1569, and to relate the presence of a young star cluster located in the centre of the field-of-view to the surrounding gas. This field is part of a large, bright H\two{} region complex \citep[H\two\ region 2;][]{waller91} that dominates the emission intensity, and exhibits evidence for recent, on-going and embedded star- and cluster-formation, presumably fuelled by the large near-by repository of molecular gas \citep{taylor99}.

Summing the spectra from an aperture centred on the position of the cluster has allowed us to measure and derive a number of fundamental characteristics of this object. We find:
\begin{itemize}
  \item Cluster 10 is composed of two sub-clusters, which we have denoted 10A and 10B, with sizes $R_{\rm eff, A} = 0.88$ and $R_{\rm eff, B} = 0.60$~pc and a projected separation of 3.7~pc. Because of their proximity, they cannot be resolved in our IFU data.
  \item Using model colour-colour diagrams, we were able to estimate their ages to be 5--7~Myr for 10A and $\leq$5~Myr for 10B, with reddenings consistent with the Galactic foreground level (confirmed by measurements of the H$\alpha$/H$\beta$ ratio from our IFU spectrum).
  \item We derive a photometric mass of $7 \pm 5 \times 10^{3}$~\Msol{} from the combined light of both sub-clusters. This result means that cluster 10 is not a super star cluster in the generally accepted definition. 
  \item The presence of the red WR-bump (resulting from broad C\four{} emission) confirms the young age of at least one of the sub-clusters. From measuring the total luminosity of the feature, we estimate an equivalent population of $10\pm 5$ WC4 stars in cluster 10. This is consistent with the narrow-band He\two{} observations of \citet{buckalew00} who find emission associated with 10B equating to a luminosity equivalent of 3 WNL stars.
\end{itemize}
The spatially resolved nature of our observations have allowed us to make a detailed analysis of the distribution of gas within the FoV. The most important findings are:
\begin{itemize}
  \item We find bright emission over the whole field well characterised by a Gaussian function with mildly supersonic line widths (C1), and a distinct lower limit of FWHM$_{0}\approx 35$~\kms{} (equating to $\sigma = 12$~\kms{} after correction for thermal broadening).
  \item We also measure a fainter line component (C2) with line widths of between $\sim$100--300~\kms{}, also found over the whole field.
  \item The width of the broad underlying emission is both highly correlated with the position of the bright narrow component emission (i.e.~bright C1 $\propto$ broad C2), and anti-correlated with its own flux (i.e.~bright C2 $\propto$ narrow C2).
  \item We conclude that C1 represents the turbulent ISM, where its motions result from a convolution of gravitational virial motions and the general stirring effects of the starburst (giving rise to the clear lower-limit to the line width, FWHM$_{0}$). However, it is possible that an additional minor contribution to its width may originate from unresolved shell components along the line-of-sight explaining the local variations in C1 width across the FoV.
  \item We find the most convincing mechanism to explain the broad component is emission from a turbulent mixing layer on the surface of the cool gas clumps in the ISM of NGC 1569. This layer results from the coupling between the clump material and the hot, fast winds from the surrounding clusters, and gives rise, through thermal evaporation and/or ablation, to a tenuous, highly turbulent, diffuse gas that pervades the local environment that will eventually mass-load the outflow.
  \item The distribution of non-photoionized (shocked) gas as determined from nebular diagnostic line ratios supports evidence that the dominant source of wind energy/material is from SSC A, consistent with morphological indicators such as swept-back cones around dense H$\alpha$ emitting knots that point towards this cluster. This does not mean, however, that the significant proportion of ionizing flux in this region does not come from the young cluster 10.
    \item The average radial velocity offset between C1 and C2 is very small compared to the average line-width of C2, indicating that in our FoV, turbulent motions dominate over bulk flows, and that we are sampling well within the `energy injection zone' of the outflow.
\end{itemize}

Our results reflect the presence of a complex ISM with a wide range in physical conditions. Our suggestion that gas is being mixed from the relatively cool photoionized medium into the hot phase carries with it a number of astrophysical implications that could generally apply to galactic winds. Among these would be mass loading of the hot ISM occurs close to or within the star forming disk. This will enhance the hot ISM cooling rate and increase the density of material injected into the base of the thermal wind, thus influencing the properties of the wind. A full discussion of the ionized gas state in NGC 1569 will be presented in a companion paper \nocite{westm07b}(Westmoquette et al.\ 2007a; Paper II), where we put our findings for this IFU position in context with the whole central region using observations made with three other IFU pointings. In a forthcoming contribution (Westmoquette et al. in prep.; Paper III) we will examine the properties of the galactic outflow as a whole using deep H$\alpha$ imaging and IFU observations and attempt to create a unified physical picture of the outflow state.

%%%%%%%%%%%%%%%%%%%%%%%%%%%%%%%%%%
\section*{Acknowledgements}
We thank the referee for their insightful comments and suggestions that have improved the paper.
MSW thanks the Instituto de Astrof\'isca de Canarias (IAC) for their warm hospitality and financial support during the writing of this paper. MSW also wishes to thank Roberto Terlevich for interesting discussions on the origin of the line components. KME acknowledges the support from the Euro3D Research Training Network, grant no. HORN-CT2002-00305.
The Gemini Observatory is operated by the Association of Universities for Research in Astronomy, Inc., under a cooperative agreement with the NSF on behalf of the Gemini partnership: the National Science Foundation (United States), the Particle Physics and Astronomy Research Council (United Kingdom), the National Research Council (Canada), CONICYT (Chile), the Australian Research Council (Australia), CNPq (Brazil) and CONICET (Argentina).

%%%%%%%%%%%%%%%%%%%%%%%%%%%%%%%%%%
% Bibliography
%%%%%%%%%%%%%%%%%%%%%%%%%%%%%%%%%%
\bibliographystyle{mn2e}
\bibliography{/Users/msw/Documents/work/Thesis/thesis/references}

\begin{thebibliography}{}

\bibitem[\protect\citeauthoryear{{Ables}}{{Ables}}{1971}]{ables71}
{Ables} H.~D.,  1971, Publ. U.S. Naval Obs., 20, 60

\bibitem[\protect\citeauthoryear{{Allington-Smith}, {Murray}, {Content},
  {Dodsworth}, {Davies}, {Miller}, {Jorgensen}, {Hook}
  et~al.,}{{Allington-Smith} et~al.}{2002}]{allington02}
{Allington-Smith} J.,  {Murray} G.,  {Content} R.,  {Dodsworth} G.,  {Davies}
  R.,  {Miller} B.~W.,  {Jorgensen} I.,  {Hook} I.,    et~al., 2002, \pasp,
  114, 892

\bibitem[\protect\citeauthoryear{{Anders}, {de Grijs}, {Fritze-v.~Alvensleben}
  \& {Bissantz}}{{Anders} et~al.}{2004}]{anders04}
{Anders} P.,  {de Grijs} R.,  {Fritze-v.~Alvensleben} U.,    {Bissantz} N.,
  2004, \mnras, 347, 17

\bibitem[\protect\citeauthoryear{{Arp} \& {Sandage}}{{Arp} \&
  {Sandage}}{1985}]{arp85}
{Arp} H.,  {Sandage} A.,  1985, \aj, 90, 1163

\bibitem[\protect\citeauthoryear{{Arribas}, {Mediavilla},
  {Garc{\'{\i}}a-Lorenzo}, {del Burgo} \& {Fuensalida}}{{Arribas}
  et~al.}{1999}]{arribas99}
{Arribas} S.,  {Mediavilla} E.,  {Garc{\'{\i}}a-Lorenzo} B.,  {del Burgo} C.,
   {Fuensalida} J.~J.,  1999, \aaps, 136, 189

\bibitem[\protect\citeauthoryear{{Begelman} \& {Fabian}}{{Begelman} \&
  {Fabian}}{1990}]{bf90}
{Begelman} M.~C.,  {Fabian} A.~C.,  1990, \mnras, 244, 26P

\bibitem[\protect\citeauthoryear{{Bosch}, {Selman}, {Melnick} \&
  {Terlevich}}{{Bosch} et~al.}{2001}]{bosch01}
{Bosch} G.,  {Selman} F.,  {Melnick} J.,    {Terlevich} R.,  2001, \aap, 380,
  137

\bibitem[\protect\citeauthoryear{{Bruzual} \& {Charlot}}{{Bruzual} \&
  {Charlot}}{2003}]{bc03}
{Bruzual} G.,  {Charlot} S.,  2003, \mnras, 344, 1000

\bibitem[\protect\citeauthoryear{{Buckalew}, {Dufour}, {Shopbell} \&
  {Walter}}{{Buckalew} et~al.}{2000}]{buckalew00}
{Buckalew} B.~A.,  {Dufour} R.~J.,  {Shopbell} P.~L.,    {Walter} D.~K.,  2000,
  \aj, 120, 2402

\bibitem[\protect\citeauthoryear{{Buckalew} \& {Kobulnicky}}{{Buckalew} \&
  {Kobulnicky}}{2006}]{buckalew06}
{Buckalew} B.~A.,  {Kobulnicky} H.~A.,  2006, \aj, 132, 1061

\bibitem[\protect\citeauthoryear{{Chu} \& {Kennicutt}}{{Chu} \&
  {Kennicutt}}{1994}]{chuken94}
{Chu} Y.-H.,  {Kennicutt} R.~C.,  1994, \apj, 425, 720

\bibitem[\protect\citeauthoryear{{Crane}, {Hegyi} \& {Lambert}}{{Crane}
  et~al.}{1991}]{crane91}
{Crane} P.,  {Hegyi} D.~J.,    {Lambert} D.~L.,  1991, \apj, 378, 181

\bibitem[\protect\citeauthoryear{{Crowther}, {De Marco} \& {Barlow}}{{Crowther}
  et~al.}{1998}]{crowther98}
{Crowther} P.~A.,  {De Marco} O.,    {Barlow} M.~J.,  1998, \mnras, 296, 367

\bibitem[\protect\citeauthoryear{{Crowther} \& {Hadfield}}{{Crowther} \&
  {Hadfield}}{2006}]{crowther06}
{Crowther} P.~A.,  {Hadfield} L.~J.,  2006, \aap, 449, 711

\bibitem[\protect\citeauthoryear{{de Marchi}, {Clampin}, {Greggio},
  {Leitherer}, {Nota} \& {Tosi}}{{de Marchi} et~al.}{1997}]{demarchi97}
{de Marchi} G.,  {Clampin} M.,  {Greggio} L.,  {Leitherer} C.,  {Nota} A.,
  {Tosi} M.,  1997, \apjl, 479, L27

\bibitem[\protect\citeauthoryear{{de Vaucouleurs}, {de Vaucouleurs}, {Corwin}
  Jr., {Buta}, {Paturel} \& {Fouque}}{{de Vaucouleurs}
  et~al.}{1991}]{vaucouleurs91}
{de Vaucouleurs} G.,  {de Vaucouleurs} A.,  {Corwin} Jr. H.~G.,  {Buta} R.~J.,
  {Paturel} G.,    {Fouque} P.,  1991, {Third Reference Catalogue of Bright
  Galaxies}.
Volume 1-3, XII, 2069~Springer-Verlag Berlin Heidelberg New York

\bibitem[\protect\citeauthoryear{{Devine} \& {Bally}}{{Devine} \&
  {Bally}}{1999}]{db99}
{Devine} D.,  {Bally} J.,  1999, \apj, 510, 197

\bibitem[\protect\citeauthoryear{{Devost}, {Roy} \& {Drissen}}{{Devost}
  et~al.}{1997}]{devost97}
{Devost} D.,  {Roy} J.-R.,    {Drissen} L.,  1997, \apj, 482, 765

\bibitem[\protect\citeauthoryear{{Dimeo}}{{Dimeo}}{2005}]{dimeo}
{Dimeo} R.,  2005, PAN User Guide

\bibitem[\protect\citeauthoryear{{Dopita}, {Kewley}, {Heisler} \&
  {Sutherland}}{{Dopita} et~al.}{2000}]{dopita00}
{Dopita} M.~A.,  {Kewley} L.~J.,  {Heisler} C.~A.,    {Sutherland} R.~S.,
  2000, \apj, 542, 224

\bibitem[\protect\citeauthoryear{{Dyson}}{{Dyson}}{1979}]{dyson79}
{Dyson} J.~E.,  1979, \aap, 73, 132

\bibitem[\protect\citeauthoryear{{Dyson}, {Hartquist}, {Malone} \&
  {Taylor}}{{Dyson} et~al.}{1995}]{dyson95}
{Dyson} J.~E.,  {Hartquist} T.~W.,  {Malone} M.~T.,    {Taylor} S.~D.,  1995,
  in {Lizano} S.,  {Torrelles} J.~M.,  eds, Revista Mexicana de Astronomia y
  Astrofisica Conference Series {Boundary Layers and Highly Supersonic
  Molecular Hydrogen Flows}.
p.~119

\bibitem[\protect\citeauthoryear{{Falgarone} \& {Phillips}}{{Falgarone} \&
  {Phillips}}{1990}]{falgarone90}
{Falgarone} E.,  {Phillips} T.~G.,  1990, \apj, 359, 344

\bibitem[\protect\citeauthoryear{{Fruchter} \& {Hook}}{{Fruchter} \&
  {Hook}}{2002}]{fruchter02}
{Fruchter} A.~S.,  {Hook} R.~N.,  2002, \pasp, 114, 144

\bibitem[\protect\citeauthoryear{{Galliano}, {Madden}, {Jones}, {Wilson},
  {Bernard} \& {Le Peintre}}{{Galliano} et~al.}{2003}]{galliano03}
{Galliano} F.,  {Madden} S.~C.,  {Jones} A.~P.,  {Wilson} C.~D.,  {Bernard}
  J.-P.,    {Le Peintre} F.,  2003, \aap, 407, 159

\bibitem[\protect\citeauthoryear{{Geballe}, {Persson}, {McGregor}, {Simon} \&
  {Lonsdale}}{{Geballe} et~al.}{1986}]{geballe86}
{Geballe} T.~R.,  {Persson} S.~E.,  {McGregor} P.~J.,  {Simon} T.,
  {Lonsdale} C.~J.,  1986, \apj, 302, 693

\bibitem[\protect\citeauthoryear{{Gonz\'alez-Delgado}, {Leitherer}, {Heckman}
  \& {Cervi{\~n}o}}{{Gonz\'alez-Delgado} et~al.}{1997}]{g-d97}
{Gonz\'alez-Delgado} R.~M.,  {Leitherer} C.,  {Heckman} T.,    {Cervi{\~n}o}
  M.,  1997, \apj, 483, 705

\bibitem[\protect\citeauthoryear{{Gonz\'alez-Delgado}, {Perez},
  {Tenorio-Tagle}, {Vilchez}, {Terlevich}, {Terlevich}, {Telles},
  {Rodriguez-Espinosa} et~al.,}{{Gonz\'alez-Delgado} et~al.}{1994}]{g-d94}
{Gonz\'alez-Delgado} R.~M.,  {Perez} E.,  {Tenorio-Tagle} G.,  {Vilchez} J.~M.,
   {Terlevich} E.,  {Terlevich} R.,  {Telles} E.,  {Rodriguez-Espinosa} J.~M.,
    et~al., 1994, \apj, 437, 239

\bibitem[\protect\citeauthoryear{{Greggio}, {Tosi}, {Clampin}, {de Marchi},
  {Leitherer}, {Nota} \& {Sirianni}}{{Greggio} et~al.}{1998}]{greggio98}
{Greggio} L.,  {Tosi} M.,  {Clampin} M.,  {de Marchi} G.,  {Leitherer} C.,
  {Nota} A.,    {Sirianni} M.,  1998, \apj, 504, 725

\bibitem[\protect\citeauthoryear{{Greve}, {Tarchi}, {H{\"u}ttemeister}, {de
  Grijs}, {van der Hulst}, {Garrington} \& {Neininger}}{{Greve}
  et~al.}{2002}]{greve02}
{Greve} A.,  {Tarchi} A.,  {H{\"u}ttemeister} S.,  {de Grijs} R.,  {van der
  Hulst} J.~M.,  {Garrington} S.~T.,    {Neininger} N.,  2002, \aap, 381, 825

\bibitem[\protect\citeauthoryear{{Hartquist}, {Dyson} \&
  {Williams}}{{Hartquist} et~al.}{1992}]{hartquist92}
{Hartquist} T.~W.,  {Dyson} J.~E.,    {Williams} D.~A.,  1992, \mnras, 257, 419

\bibitem[\protect\citeauthoryear{{Heckman}, {Dahlem}, {Lehnert}, {Fabbiano},
  {Gilmore} \& {Waller}}{{Heckman} et~al.}{1995}]{heckman95}
{Heckman} T.~M.,  {Dahlem} M.,  {Lehnert} M.~D.,  {Fabbiano} G.,  {Gilmore} D.,
     {Waller} W.~H.,  1995, \apj, 448, 98

\bibitem[\protect\citeauthoryear{{Heckman} \& {Lehnert}}{{Heckman} \&
  {Lehnert}}{2000}]{heckman00b}
{Heckman} T.~M.,  {Lehnert} M.~D.,  2000, \apj, 537, 690

\bibitem[\protect\citeauthoryear{{Heckman}, {Sembach}, {Meurer}, {Strickland},
  {Martin}, {Calzetti} \& {Leitherer}}{{Heckman} et~al.}{2001}]{heckman01}
{Heckman} T.~M.,  {Sembach} K.~R.,  {Meurer} G.~R.,  {Strickland} D.~K.,
  {Martin} C.~L.,  {Calzetti} D.,    {Leitherer} C.,  2001, \apj, 554, 1021

\bibitem[\protect\citeauthoryear{{Herbig}}{{Herbig}}{1995}]{herbig95}
{Herbig} G.~H.,  1995, \araa, 33, 19

\bibitem[\protect\citeauthoryear{{Holtzman}, {Burrows}, {Casertano}, {Hester},
  {Trauger}, {Watson} \& {Worthey}}{{Holtzman} et~al.}{1995}]{holtzman95}
{Holtzman} J.~A.,  {Burrows} C.~J.,  {Casertano} S.,  {Hester} J.~J.,
  {Trauger} J.~T.,  {Watson} A.~M.,    {Worthey} G.,  1995, \pasp, 107, 1065

\bibitem[\protect\citeauthoryear{{Homeier} \& {Gallagher}}{{Homeier} \&
  {Gallagher}}{1999}]{homeier99}
{Homeier} N.~L.,  {Gallagher} J.~S.,  1999, \apj, 522, 199

\bibitem[\protect\citeauthoryear{{Howarth}}{{Howarth}}{1983}]{howarth83}
{Howarth} I.~D.,  1983, \mnras, 203, 301

\bibitem[\protect\citeauthoryear{{Hummer} \& {Storey}}{{Hummer} \&
  {Storey}}{1987}]{humstor87}
{Hummer} D.~G.,  {Storey} P.~J.,  1987, \mnras, 224, 801

\bibitem[\protect\citeauthoryear{{Hunter}, {Hawley} \& {Gallagher}}{{Hunter}
  et~al.}{1993}]{hunter93}
{Hunter} D.~A.,  {Hawley} W.~N.,    {Gallagher} J.~S.,  1993, \aj, 106, 1797

\bibitem[\protect\citeauthoryear{{Hunter}, {O'Connell}, {Gallagher} \&
  {Smecker-Hane}}{{Hunter} et~al.}{2000}]{hunter00}
{Hunter} D.~A.,  {O'Connell} R.~W.,  {Gallagher} J.~S.,    {Smecker-Hane}
  T.~A.,  2000, \aj, 120, 2383

\bibitem[\protect\citeauthoryear{{Hunter}, {Shaya}, {Holtzman}, {Light},
  {O'Neil} Jr. \& {Lynds}}{{Hunter} et~al.}{1995}]{hunter95}
{Hunter} D.~A.,  {Shaya} E.~J.,  {Holtzman} J.~A.,  {Light} R.~M.,  {O'Neil}
  Jr. E.~J.,    {Lynds} R.,  1995, \apj, 448, 179

\bibitem[\protect\citeauthoryear{{Israel}}{{Israel}}{1988}]{israel88}
{Israel} F.~P.,  1988, \aap, 194, 24

\bibitem[\protect\citeauthoryear{{Izotov}, {Dyak}, {Chaffee}, {Foltz},
  {Kniazev} \& {Lipovetsky}}{{Izotov} et~al.}{1996}]{izotov96}
{Izotov} Y.~I.,  {Dyak} A.~B.,  {Chaffee} F.~H.,  {Foltz} C.~B.,  {Kniazev}
  A.~Y.,    {Lipovetsky} V.~A.,  1996, \apj, 458, 524

\bibitem[\protect\citeauthoryear{{Kewley}, {Dopita}, {Sutherland}, {Heisler} \&
  {Trevena}}{{Kewley} et~al.}{2001}]{kewley01}
{Kewley} L.~J.,  {Dopita} M.~A.,  {Sutherland} R.~S.,  {Heisler} C.~A.,
  {Trevena} J.,  2001, \apj, 556, 121

\bibitem[\protect\citeauthoryear{{Klein}, {McKee} \& {Colella}}{{Klein}
  et~al.}{1994}]{klein94}
{Klein} R.~I.,  {McKee} C.~F.,    {Colella} P.,  1994, \apj, 420, 213

\bibitem[\protect\citeauthoryear{{Kobulnicky} \& {Skillman}}{{Kobulnicky} \&
  {Skillman}}{1997}]{kobulnicky97}
{Kobulnicky} H.~A.,  {Skillman} E.~D.,  1997, \apj, 489, 636

\bibitem[\protect\citeauthoryear{{Krist}}{{Krist}}{2004}]{krist04}
{Krist} J.,  2004, The Tiny Tim User's Guide v.\ 6.3

\bibitem[\protect\citeauthoryear{{Larsen}}{{Larsen}}{1999}]{larsen99b}
{Larsen} S.~S.,  1999, \aaps, 139, 393

\bibitem[\protect\citeauthoryear{{Leitherer}, {Schaerer}, {Goldader},
  {Delgado}, {Robert}, {Kune}, {de Mello}, {Devost} et~al.,}{{Leitherer}
  et~al.}{1999}]{leitherer99}
{Leitherer} C.,  {Schaerer} D.,  {Goldader} J.~D.,  {Delgado} R.~M.~G.,
  {Robert} C.,  {Kune} D.~F.,  {de Mello} D.~F.,  {Devost} D.,    et~al., 1999,
  \apjs, 123, 3

\bibitem[\protect\citeauthoryear{{Lisenfeld}, {Israel}, {Stil} \&
  {Sievers}}{{Lisenfeld} et~al.}{2002}]{lisenfeld02}
{Lisenfeld} U.,  {Israel} F.~P.,  {Stil} J.~M.,    {Sievers} A.,  2002, \aap,
  382, 860

\bibitem[\protect\citeauthoryear{{Lisenfeld}, {Wilding}, {Pooley} \&
  {Alexander}}{{Lisenfeld} et~al.}{2004}]{lisenfeld04}
{Lisenfeld} U.,  {Wilding} T.~W.,  {Pooley} G.~G.,    {Alexander} P.,  2004,
  \mnras, 349, 1335

\bibitem[\protect\citeauthoryear{{Maoz}, {Ho} \& {Sternberg}}{{Maoz}
  et~al.}{2001}]{maoz01}
{Maoz} D.,  {Ho} L.~C.,    {Sternberg} A.,  2001, \apjl, 554, L139

\bibitem[\protect\citeauthoryear{{Marcolini}, {Strickland}, {D'Ercole},
  {Heckman} \& {Hoopes}}{{Marcolini} et~al.}{2005}]{marcolini05}
{Marcolini} A.,  {Strickland} D.~K.,  {D'Ercole} A.,  {Heckman} T.~M.,
  {Hoopes} C.~G.,  2005, \mnras, 362, 626

\bibitem[\protect\citeauthoryear{{Marlowe}, {Heckman}, {Wyse} \&
  {Schommer}}{{Marlowe} et~al.}{1995}]{marlowe95}
{Marlowe} A.~T.,  {Heckman} T.~M.,  {Wyse} R.~F.~G.,    {Schommer} R.,  1995,
  \apj, 438, 563

\bibitem[\protect\citeauthoryear{{Martin}}{{Martin}}{1998}]{martin98}
{Martin} C.~L.,  1998, \apj, 506, 222

\bibitem[\protect\citeauthoryear{{Martin}, {Kobulnicky} \& {Heckman}}{{Martin}
  et~al.}{2002}]{martin02}
{Martin} C.~L.,  {Kobulnicky} H.~A.,    {Heckman} T.~M.,  2002, \apj, 574, 663

\bibitem[\protect\citeauthoryear{{Martinez-Delgado}, {Tenorio-Tagle},
  {Munoz-Tunon}, {Moiseev} \& {Cairos}}{{Martinez-Delgado}
  et~al.}{2007}]{martinez07}
{Martinez-Delgado} I.,  {Tenorio-Tagle} G.,  {Munoz-Tunon} C.,  {Moiseev}
  A.~V.,    {Cairos} L.~M.,  2007, astro-ph/0703165

\bibitem[\protect\citeauthoryear{{Melnick}}{{Melnick}}{1977}]{melnick77}
{Melnick} J.,  1977, \apj, 213, 15

\bibitem[\protect\citeauthoryear{{Melnick}, {Tenorio-Tagle} \&
  {Terlevich}}{{Melnick} et~al.}{1999}]{melnick99}
{Melnick} J.,  {Tenorio-Tagle} G.,    {Terlevich} R.,  1999, \mnras, 302, 677

\bibitem[\protect\citeauthoryear{{Melnick}, {Terlevich} \& {Moles}}{{Melnick}
  et~al.}{1988}]{melnick88}
{Melnick} J.,  {Terlevich} R.,    {Moles} M.,  1988, \mnras, 235, 297

\bibitem[\protect\citeauthoryear{{Mu{\~n}oz-Tu{\~n}{\'o}n}, {Tenorio-Tagle},
  {Castaneda} \& {Terlevich}}{{Mu{\~n}oz-Tu{\~n}{\'o}n} et~al.}{1996}]{m-t96}
{Mu{\~n}oz-Tu{\~n}{\'o}n} C.,  {Tenorio-Tagle} G.,  {Castaneda} H.~O.,
  {Terlevich} R.,  1996, \aj, 112, 1636

\bibitem[\protect\citeauthoryear{{M{\"u}hle}, {Klein}, {Wilcots} \&
  {H{\"u}ttemeister}}{{M{\"u}hle} et~al.}{2005}]{muhle05}
{M{\"u}hle} S.,  {Klein} U.,  {Wilcots} E.~M.,    {H{\"u}ttemeister} S.,  2005,
  \aj, 130, 524

\bibitem[\protect\citeauthoryear{{Oke}}{{Oke}}{1990}]{oke90}
{Oke} J.~B.,  1990, \aj, 99, 1621

\bibitem[\protect\citeauthoryear{{Origlia}, {Leitherer}, {Aloisi}, {Greggio} \&
  {Tosi}}{{Origlia} et~al.}{2001}]{origlia01}
{Origlia} L.,  {Leitherer} C.,  {Aloisi} A.,  {Greggio} L.,    {Tosi} M.,
  2001, \aj, 122, 815

\bibitem[\protect\citeauthoryear{{Osterbrock}}{{Osterbrock}}{1989}]{osterbrock%
89}
{Osterbrock} D.~E.,  1989, Astrophysics of Gaseous Nebulae and Active Galactic
  Nuclei.
University Science Books

\bibitem[\protect\citeauthoryear{{Pittard}}{{Pittard}}{2006}]{pittard06}
{Pittard} J.~M.,  2006, astro-ph/0607310

\bibitem[\protect\citeauthoryear{{Pittard}, {Dyson}, {Falle} \&
  {Hartquist}}{{Pittard} et~al.}{2005}]{pittard05}
{Pittard} J.~M.,  {Dyson} J.~E.,  {Falle} S.~A.~E.~G.,    {Hartquist} T.~W.,
  2005, \mnras, 361, 1077

\bibitem[\protect\citeauthoryear{{Reakes}}{{Reakes}}{1980}]{reakes80}
{Reakes} M.,  1980, \mnras, 192, 297

\bibitem[\protect\citeauthoryear{{Rela{\~n}o}, {Lisenfeld}, {Vilchez} \&
  {Battaner}}{{Rela{\~n}o} et~al.}{2006}]{relano06}
{Rela{\~n}o} M.,  {Lisenfeld} U.,  {Vilchez} J.~M.,    {Battaner} E.,  2006,
  \aap, 452, 413

\bibitem[\protect\citeauthoryear{{Roy}, {Aube}, {McCall} \& {Dufour}}{{Roy}
  et~al.}{1992}]{roy92}
{Roy} J.-R.,  {Aube} M.,  {McCall} M.~L.,    {Dufour} R.~J.,  1992, \apj, 386,
  498

\bibitem[\protect\citeauthoryear{{S{\'a}nchez}}{{S{\'a}nchez}}{2004}]{sanchez0%
4}
{S{\'a}nchez} S.~F.,  2004, Astronomische Nachrichten, 325, 167

\bibitem[\protect\citeauthoryear{{Scalo}}{{Scalo}}{1987}]{scalo87}
{Scalo} J.~M.,  1987, in {Hollenbach} D.~J.,  {Thronson} Jr. H.~A.,  eds, ASSL
  Vol. 134: Interstellar Processes {Theoretical approaches to interstellar
  turbulence}.
pp 349--392

\bibitem[\protect\citeauthoryear{{Shopbell} \& {Bland-Hawthorn}}{{Shopbell} \&
  {Bland-Hawthorn}}{1998}]{shopbell98}
{Shopbell} P.~L.,  {Bland-Hawthorn} J.,  1998, \apj, 493, 129

\bibitem[\protect\citeauthoryear{{Sidoli}, {Smith} \& {Crowther}}{{Sidoli}
  et~al.}{2006}]{sidoli06}
{Sidoli} F.,  {Smith} L.~J.,    {Crowther} P.~A.,  2006, \mnras, 370, 799

\bibitem[\protect\citeauthoryear{{Sirianni}, {Jee}, {Ben{\'{\i}}tez},
  {Blakeslee}, {Martel}, {Meurer}, {Clampin}, {De Marchi}, {Ford}, {Gilliland},
  {Hartig}, {Illingworth}, {Mack} \& {McCann}}{{Sirianni}
  et~al.}{2005}]{sirianni05}
{Sirianni} M.,  {Jee} M.~J.,  {Ben{\'{\i}}tez} N.,  {Blakeslee} J.~P.,
  {Martel} A.~R.,  {Meurer} G.,  {Clampin} M.,  {De Marchi} G.,  {Ford} H.~C.,
  {Gilliland} R.,  {Hartig} G.~F.,  {Illingworth} G.~D.,  {Mack} J.,
  {McCann} W.~J.,  2005, \pasp, 117, 1049

\bibitem[\protect\citeauthoryear{{Slavin}, {Shull} \& {Begelman}}{{Slavin}
  et~al.}{1993}]{slavin93}
{Slavin} J.~D.,  {Shull} J.~M.,    {Begelman} M.~C.,  1993, \apj, 407, 83

\bibitem[\protect\citeauthoryear{{Stil} \& {Israel}}{{Stil} \&
  {Israel}}{1998}]{stil98}
{Stil} J.~M.,  {Israel} F.~P.,  1998, \aap, 337, 64

\bibitem[\protect\citeauthoryear{{Stil} \& {Israel}}{{Stil} \&
  {Israel}}{2002}]{stil02}
{Stil} J.~M.,  {Israel} F.~P.,  2002, \aap, 392, 473

\bibitem[\protect\citeauthoryear{{Storey} \& {Hummer}}{{Storey} \&
  {Hummer}}{1995}]{storey95}
{Storey} P.~J.,  {Hummer} D.~G.,  1995, \mnras, 272, 41

\bibitem[\protect\citeauthoryear{{Suchkov}, {Berman}, {Heckman} \&
  {Balsara}}{{Suchkov} et~al.}{1996}]{suchkov96}
{Suchkov} A.~A.,  {Berman} V.~G.,  {Heckman} T.~M.,    {Balsara} D.~S.,  1996,
  \apj, 463, 528

\bibitem[\protect\citeauthoryear{{Taylor}, {H{\"u}ttemeister}, {Klein} \&
  {Greve}}{{Taylor} et~al.}{1999}]{taylor99}
{Taylor} C.~L.,  {H{\"u}ttemeister} S.,  {Klein} U.,    {Greve} A.,  1999,
  \aap, 349, 424

\bibitem[\protect\citeauthoryear{{Tenorio-Tagle}, {Mu{\~n}oz-Tu{\~n}{\'o}n} \&
  {Cid-Fernandes}}{{Tenorio-Tagle} et~al.}{1996}]{t-t96}
{Tenorio-Tagle} G.,  {Mu{\~n}oz-Tu{\~n}{\'o}n} C.,    {Cid-Fernandes} R.,
  1996, \apj, 456, 264

\bibitem[\protect\citeauthoryear{{Tenorio-Tagle}, {Mu{\~n}oz-Tu{\~n}{\'o}n} \&
  {Cox}}{{Tenorio-Tagle} et~al.}{1993}]{t-t93}
{Tenorio-Tagle} G.,  {Mu{\~n}oz-Tu{\~n}{\'o}n} C.,    {Cox} D.~P.,  1993, \apj,
  418, 767

\bibitem[\protect\citeauthoryear{{Tenorio-Tagle}, {Mu{\~n}oz-Tu{\~n}{\'o}n},
  {P{\'e}rez}, {Silich} \& {Telles}}{{Tenorio-Tagle} et~al.}{2006}]{t-t06}
{Tenorio-Tagle} G.,  {Mu{\~n}oz-Tu{\~n}{\'o}n} C.,  {P{\'e}rez} E.,  {Silich}
  S.,    {Telles} E.,  2006, \apj, 643, 186

\bibitem[\protect\citeauthoryear{{Terlevich} \& {Melnick}}{{Terlevich} \&
  {Melnick}}{1981}]{terlevich81}
{Terlevich} R.,  {Melnick} J.,  1981, \mnras, 195, 839

\bibitem[\protect\citeauthoryear{{Tokura}, {Onaka}, {Takahashi}, {Miyata},
  {Sako}, {Honda}, {Okada}, {Sakon} et~al.,}{{Tokura} et~al.}{2006}]{tokura06}
{Tokura} D.,  {Onaka} T.,  {Takahashi} H.,  {Miyata} T.,  {Sako} S.,  {Honda}
  M.,  {Okada} Y.,  {Sakon} I.,    et~al., 2006, \apj, 648, 355

\bibitem[\protect\citeauthoryear{{Tomita}, {Ohta} \& {Saito}}{{Tomita}
  et~al.}{1994}]{tomita94}
{Tomita} A.,  {Ohta} K.,    {Saito} M.,  1994, \pasj, 46, 335

\bibitem[\protect\citeauthoryear{{Vacca} \& {Conti}}{{Vacca} \&
  {Conti}}{1992}]{vacca92}
{Vacca} W.~D.,  {Conti} P.~S.,  1992, \apj, 401, 543

\bibitem[\protect\citeauthoryear{{van Dokkum}}{{van
  Dokkum}}{2001}]{vandokkum01}
{van Dokkum} P.~G.,  2001, \pasp, 113, 1420

\bibitem[\protect\citeauthoryear{{Veilleux} \& {Osterbrock}}{{Veilleux} \&
  {Osterbrock}}{1987}]{veilleux87}
{Veilleux} S.,  {Osterbrock} D.~E.,  1987, \apjs, 63, 295

\bibitem[\protect\citeauthoryear{{Viallefond}, {Allen} \& {Goss}}{{Viallefond}
  et~al.}{1981}]{viallefond81}
{Viallefond} F.,  {Allen} R.~J.,    {Goss} W.~M.,  1981, \aap, 104, 127

\bibitem[\protect\citeauthoryear{{Waller}}{{Waller}}{1991}]{waller91}
{Waller} W.~H.,  1991, \apj, 370, 144

\bibitem[\protect\citeauthoryear{{Westmoquette}, {Smith}, {Gallagher} III \&
  {Exter}}{{Westmoquette} et~al.}{2007a}]{westm07b}
{Westmoquette} M.~S.,  {Smith} L.~J.,  {Gallagher} III J.~S.,    {Exter} K.~M.,
   2007a, \mnras, in press, Paper II

\bibitem[\protect\citeauthoryear{{Westmoquette}, {Smith}, {Gallagher} III,
  {O'Connell}, {Rosario} \& {de Grijs}}{{Westmoquette} et~al.}{2007b}]{westm07c}
{Westmoquette} M.~S.,  {Smith} L.~J.,  {Gallagher} III J.~S.,  {O'Connell}
  R.~W.,  {Rosario} D.~J.,    {de Grijs} R.,  2007b, \apj, in press

\bibitem[\protect\citeauthoryear{{Yang}, {Chu}, {Skillman} \&
  {Terlevich}}{{Yang} et~al.}{1996}]{yang96}
{Yang} H.,  {Chu} Y.-H.,  {Skillman} E.~D.,    {Terlevich} R.,  1996, \aj, 112,
  146

\end{thebibliography}
\bsp
\label{lastpage}

\end{document}